\documentclass[12pt]{article}
\usepackage{graphics}
\usepackage[dvips]{epsfig}

\def\lap{\mathrel{\mathpalette\fun <}}
\def\gap{\mathrel{\mathpalette\fun >}}
\def\fun#1#2{\lower3.6pt\vbox{\baselineskip0pt\lineskip.9pt
  \ialign{$\mathsurround=0pt#1\hfil##\hfil$\crcr#2\crcr\sim\crcr}}}

\begin{document}

\title{A New Method for Determining Cumulative Gravitational Lensing
Effects in Inhomogeneous Universes} 
\author{Daniel E. Holz and Robert M. Wald\\ 
{\it Enrico Fermi Institute and Department of Physics}\\ 
{\it University of Chicago}\\
{\it 5640 S. Ellis Avenue}\\ 
{\it Chicago, Illinois 60637-1433}}
\maketitle

\begin{abstract}

We present a new approach to calculating the statistical distributions
for magnification, shear, and rotation of images of cosmological
sources due to gravitational lensing. In this approach one specifies
an underlying Robertson-Walker cosmological model together with
relevant information on the clumping of matter on scales much smaller
than the Hubble radius. The geodesic deviation equation is then
integrated backwards in time until the desired redshift is reached,
using a Monte Carlo procedure wherein each photon beam in effect
``creates its own universe'' as it propagates. The approach is
somewhat similar to that used in ``Swiss cheese'' models, but the
``cheese'' has been completely eliminated, the matter distribution in
the ``voids'' need not be spherically symmetric, the total mass in
each void need equal the corresponding Robertson-Walker mass only on
average, and we do not impose an ``opaque radius'' cutoff. The case
where the matter in the universe consists of point masses is studied
in detail, and it is shown that the statistical distributions of the
lensing images are essentially independent of both the mass spectrum
and the clustering properties of the point masses, provided that the
clustering is spherical. Detailed results for the distribution of the
magnification of images are presented for the point mass case, as well
as a number of other matter distributions. We apply our results (i) to
argue that the positive correlation recently found between quasar
luminosity and the number of absorption line systems is not likely to
be due to lensing, and (ii) to determine the amount of ``noise'' and
possible bias produced by lensing in measurements of $q_0$ using
distant supernovae.

\end{abstract}

\section{Introduction}
\label{intro}
In recent years there has been a great deal of interest in studying
the effects on cosmologically distant sources produced by
gravitational lensing due to intervening matter. In many cases of
interest, the lensing effects can be assumed to be produced by a
single galaxy or cluster of galaxies, and one can use the detailed
structure of the images produced by lensing to extract a great deal of
information about the mass distribution of the galaxy or cluster.
However, in other circumstances of interest one may be
interested in the cumulative lensing effects produced by many
different objects (or voids), and one may be primarily interested in
statistical distributions of the image brightenings and/or distortions,
rather than the detailed modeling of any individual lens system.

Two examples of the latter circumstances are the following: (1) Vanden
Berk {\em et al.}~\cite{vandenberk} have presented evidence for a
positive correlation between quasar luminosity and the number of
intervening Carbon IV absorption clouds. Could this correlation be the
result of the cumulative gravitational lensing effects produced by the
mass distributions associated with these clouds? (2) Efforts are
currently underway to use supernovae occurring at cosmological
distances as standard candles for tests of
$q_0$~\cite{perlmutter}. How much ``noise'' in the apparent luminosity
distribution of the supernovae would be expected from gravitational
lensing effects? Could any useful information about the distribution
of matter in the universe be extractable from this ``noise''?

The main purpose of this paper is to present a new approach for
determining cumulative gravitational lensing effects on cosmological
scales due to inhomogeneities in the matter distribution of the
universe. As explained further below, in this approach one specifies
an underlying Robertson-Walker cosmological model together with one's
assumptions concerning the detailed clumping and clustering of matter
in the universe. Both the Robertson-Walker model and the
clumping/clustering of matter may be specified arbitrarily, provided
that the clustering of matter occurs only on scales much smaller than
the Hubble radius and that the average density of the matter
distribution corresponds to that of the underlying Robertson-Walker
model. Our approach then enables one to accurately obtain statistical
distributions for the luminosity, shear, and rotation of images of
``standard candle'' (nearly) point sources at any cosmological
redshift.  When multiple images occur, however, even statistical
information about the number of images and the relationships between
the images cannot be easily extracted using our approach, since that
would require us to keep track of the relationship between finitely
(as opposed to infinitesimally) separated null
geodesics. Nevertheless, statistical information about the luminosity,
shear, and rotation of the individual images occurring in multiple
images is included in our distributions.

The rest of this section will be devoted to an overview of our approach
for determining statistical lensing effects in inhomogeneous universes.
Subsection~\ref{Cosmological Model} introduces our cosmological
model, presenting and justifying
the metric which provides the framework for our results.
Subsection~\ref{Propagation of Photon Beams} discusses lensing
effects on the propogation
of photon beams within the cosmology, while Subsection~\ref{The Local
Nature} discusses the local nature of these effects. Subsection~\ref{Our
Method} gives a general overview of our method, and
Subsection~\ref{ss:relevant_scales} discusses the relevant
scales of the model.
In Section~\ref{method} we present our procedure for calculating
statistical lensing effects in more explicit detail. In
Section~\ref{points_and_clustering} we analyze the case where all
of the matter in the universe can be treated as being comprised of
point masses (satisfying Eq.~(\ref{mlim})).  Other distributions of
mass are considered in Subsection~\ref{ss:cases}, and then in
Subsection~\ref{ss:checks} we perform some consistency checks on our
results. Applications of our work to the analysis of lensing effects
by quasar absorption systems are given in Section~\ref{york}, and
applications to the effects of lensing on supernovae luminosity
are given in Section~\ref{supernovae}.

\subsection{Cosmological Model}
\label{Cosmological Model}

To explain our approach, we first need to state our cosmological
assumptions with more precision. We assume that the spacetime metric
of the universe is globally well approximated (on {\em all} scales) by
a ``Newtonianly perturbed Robertson-Walker metric'' of the form
\begin{equation} 
ds^2 = -(1 + 2\phi)\,d \tau^2 + (1 - 2\phi) a^2(\tau)
\left[\frac{dr^2}{1-kr^2} +
r^2 (d \theta^2 + sin^2 \theta\,d \varphi^2)\right],
\label{metric} 
\end{equation}
where $k = 0, \pm 1$.  We shall refer to the metric obtained by
setting $\phi = 0$ in Eq.~(\ref{metric}) as the {\em underlying
Robertson-Walker model}.
The spatial metric of this underlying Robertson-Walker model is
$a^2 h_{ab}$, where
\begin{equation} 
h_{ab} \equiv \frac{1}{1-kr^2}\,dr_a dr_b + r^2 (d \theta_a d \theta_b +
sin^2 \theta\,d \varphi_a d \varphi_b)
\label{smetric} 
\end{equation}
is either the metric of a unit 3-sphere ($k = 1$), a unit 3-hyperboloid
($k=-1$), or flat 3-space ($k = 0$). 

Without loss of generality, we may assume that the spatial average of
$\phi$ vanishes, since a spatially constant part of $\phi$ could be
absorbed into the definitions of $\tau$ and $a$. We also assume
that throughout spacetime---or at least out to distance scales of
order $R_H$, where $R_H \equiv H^{-1} = a/\dot{a}$ denotes the Hubble
radius of the underlying Robertson-Walker model---we have
\begin{equation} 
|\phi| \ll 1 .
\label{d1phi} 
\end{equation}
We further assume that time derivatives of $\phi$ are much smaller than
spatial derivatives, i.e.,
\begin{equation} 
|\partial \phi/ \partial \tau|^2 \ll a^{-2} h^{ab} D_a \phi D_b \phi,
\label{d2phi} 
\end{equation}
with similar relations holding for the higher time derivatives. Here
$D_a$ denotes the spatial derivative operator associated with $h_{ab}$,
and $h^{ab}$ denotes the inverse of $h_{ab}$ (so $a^{-2} h^{ab}$ is
the inverse spatial metric of the underlying Robertson-Walker
model). It is important to note that spatial derivatives of $\phi$ may
locally be very large compared with scales set by the underlying
Robertson-Walker model. However, we assume that products of first
spatial derivatives of $\phi$ are small compared with second
derivatives, i.e.,
\begin{equation} 
(h^{ab} D_a \phi D_b \phi)^2 \ll h^{ac} h^{bd} D_a D_b \phi D_c D_d \phi.
\label{d3phi} 
\end{equation}
Finally, we assume that the matter stress-energy tensor, $T_{ab}$
({\em not} including the cosmological constant term), is everywhere such
that, in the rest frame of the underlying Robertson-Walker model, the
energy density of matter greatly dominates the other components of
$T_{ab}$. In this case $T_{ab}$ is approximately of the ``matter
dominated'' form
\begin{equation} 
T_{ab} \approx \rho u_a u_b,
\label{Tab} 
\end{equation}
where $u^a$ is the unit (in the metric of Eq.~(\ref{metric})) timelike
vector field orthogonal to the surfaces of constant $\tau$. 
Eqs.~(\ref{d1phi})--(\ref{Tab}) are the only assumptions we shall need
to obtain eqs.~(\ref{ee1'})--(\ref{poisson}) below.\footnote{
E. Linder (private communication) has claimed that
the approximation $\epsilon^2/\kappa \ll 1$ of references
\cite{linder}, \cite{futamase}, and \cite{futamase2}
is also needed for the validity
of our equations below. We do not agree with this claim.}
However, in Subsection~\ref{The Local Nature}
we shall also assume that there is a (co-moving) scale
${\cal R} \ll R_H$ such that no strong correlations in the density
of matter occur on scales greater than ${\cal R}$.

We now substitute the metric form of Eq.~(\ref{metric}) and the
matter stress-energy of Eq.~(\ref{Tab}) into Einstein's
equation, possibly with a nonvanishing cosmological constant,
$\Lambda$.
We make
the approximations of {Eqs.~(\ref{d1phi})--(\ref{Tab})}, and also
drop all terms (like $\Lambda \phi$ and $\rho \phi$) which are
small compared with the curvature of the underlying
Robertson-Walker metric. The nonvanishing components of Einstein's
equation then yield\footnote{In addition to the two equations given
here---which correspond to the time-time and diagonal space-space
components of Einstein's equation---there are also contributions to
the time-space components of Einstein's equation of the form $\rho
v_a$ (where $v_a$ denotes the velocity of the matter relative to the
Hubble flow), $(\dot{a}/a) D_a \phi$, and mixed time-space derivatives
of $\phi$. These terms need not everywhere be small compared with the
curvature of the underlying Robertson-Walker metric. If only these
terms were considered, the time-space components of Einstein's
equation would yield additional equations for $\phi$ which would be
inconsistent with Eq.~(\ref{poisson}) below. This difficulty is
resolved by allowing for the presence of nonvanishing time-space
components of the metric, $g_{0 \mu}$ (with $\mu = 1,2,3$), satisfying
$|g_{0 \mu}| \ll |\phi|$. The time-space components of Einstein's
equation then become, in essence, equations which determine $g_{0
\mu}$ (see Sec.~4.4a of~\cite{wald} for further details in the
ordinary Newtonian case). However, since $g_{0 \mu}$ makes a
negligible correction to the effects calculated in this paper, we
shall ignore its presence below and, correspondingly, will not
consider the time-space components of Einstein's equation.}
\begin{eqnarray}
&3 \ddot{a}/a = \Lambda - 4 \pi \rho + a^{-2} h^{ab} D_a D_b \phi&
\label{ee1} \\
&3(\dot{a}/a)^2 = \Lambda + 8 \pi \rho -2a^{-2} h^{ab} D_a D_b \phi - 3k/a^2,&
\label{ee2} 
\end{eqnarray}
where the dots denote derivatives with respect to $\tau$.  The spatial
average of these equations yields the usual form of the matter
dominated Einstein equations for the underlying Robertson-Walker
metric, namely
\begin{eqnarray}
&3 \ddot{a}/a = \Lambda - 4 \pi \bar{\rho}&
\label{ee1'} \\
&3(\dot{a}/a)^2 = \Lambda + 8 \pi \bar{\rho} - 3k/a^2,&
\label{ee2'} 
\end{eqnarray}
where $\bar{\rho}$ denotes the spatial average of $\rho$. Subtracting
Eqs.~(\ref{ee1'}) and~(\ref{ee2'}) from Eqs.~(\ref{ee1})
and~(\ref{ee2}), we find the remaining content of Einstein's
equation is that $\phi$ satisfies the Poisson equation\footnote{
Nonlinear terms in $\phi$, such as
$a^{-2}h^{ab} D_a\phi\,D_b\phi=D_a\phi\,D^a\phi$, are neglected in
eq.~(\ref{poisson}) because they are small compared with the
term linear in $\phi$ (see eq.~(\ref{d3phi})).
On the other hand, since the spatial average of $D_aD^a\phi$
vanishes, the neglect of the spatial average of nonlinear terms
like $D_a\phi\,D^a\phi$
in eqs.~(\ref{ee1'}) and~(\ref{ee2'})
is justified as follows. We have
\begin{eqnarray*}
\int_{\rm V} \!D_a\phi\,D^a\phi\,\,d{\rm V}
&=&-\int_{\rm V} \!\phi\,D_aD^a\phi\,\,d{\rm V}\\
&=&-4\pi\int_{\rm V}\!\phi\,\delta\rho\,\,d{\rm V}\\
&=&-4\pi\int_{\rm V}\!\phi\,(\rho-\bar\rho)\,\,d{\rm V}.
\end{eqnarray*}
The integral of $\phi\,\rho$ is much less than the integral of $\rho$,
as $\phi\ll1$ and $\rho$ is nonegative. The same argument
holds for the $\phi\,\bar\rho$ term. Thus, under our assumptions,
the spatial average of
$D_a\phi\,D^a\phi$ is much less than $\bar\rho$, which
justifies dropping the former in eqs.~\ref{ee1'} and~\ref{ee2'}.
}
\begin{equation} 
a^{-2} h^{ab} D_a D_b \phi = 4 \pi \delta \rho,
\label{poisson} 
\end{equation}
where
\begin{equation} 
\delta \rho \equiv \rho - \bar{\rho}.
\label{drho} 
\end{equation}
We emphasize that it is completely consistent with our assumptions
to have, locally, $\delta \rho \gg \bar{\rho}$. It is essential
that this be allowed if Eq.~(\ref{metric}), together with
Eqs.~(\ref{d1phi})--(\ref{d3phi}), are intended as an accurate description of
our universe, since we commonly find $\delta \rho \sim 10^{30} \bar{\rho}$
in our vicinity.

Thus, in our model the matter is assumed to have an energy density
much greater than its stresses, and is assumed to move
non-relativistically with respect to the Hubble flow defined by the
underlying Robertson-Walker model. However, unlike a Robertson-Walker
model, this matter may be distributed in a very inhomogeneous manner;
in particular, as already noted, the fluctuations in the mass density
may be very large compared with the spatial average of the mass
density. Consequently, the local curvature of spacetime may differ
drastically from that of a Robertson-Walker model. Nevertheless, in
our cosmological model, the Hubble flow of the matter and the causal
structure of spacetime correspond very closely to the underlying
matter dominated Robertson-Walker model, whose mass density is equal to
the average density of matter in the universe.

It is useful to examine the form taken by the metric of Eq.~(\ref{metric}) in
a locally Minkowskian frame associated with an observer moving with
the Hubble flow, which, for convenience, we take to be located at $r =
0$. To do so we define a new radial coordinate, $R$, by
\begin{equation} 
R = a r,
\label{R} 
\end{equation}
and a new time coordinate, $T$, by
\begin{equation} 
T = \tau + {1\over2}\frac{\dot{a}}{a} R^2.
\label{T} 
\end{equation}
In these new coordinates the metric of Eq.~(\ref{metric}) takes the form
\begin{eqnarray} 
ds^2 &=& - (1 + 2 \phi - R^2 \ddot{a}/a)\,dT^2 + \left(1 - 2 \phi +
R^2\left[\left({\dot{a}}/{a}\right)^2+ k/a^2\right]\right)\,dR^2 
\nonumber \\
& & \mbox{} + (1 - 2 \phi) R^2\, d\Omega^2,
\label{n1} 
\end{eqnarray}
where we have dropped all terms of order $R^3$ and higher in distance
from the origin. Transforming to an isotropic radial coordinate, then
further transforming to the corresponding Cartesian coordinates
$X,Y,Z$, and, finally, substituting from Einstein's equations
(Eqs.~(\ref{ee1'}) and~(\ref{ee2'})) for the underlying Robertson-Walker model,
we obtain
\begin{equation} 
ds^2 = - (1 + 2 \Phi - \Lambda R^2/3)\,dT^2 + (1 - 2 \Phi - \Lambda
R^2/6)[dX^2 + dY^2 + dZ^2],
\label{n2} 
\end{equation}
where
\begin{equation} 
\Phi \equiv \phi + 2 \pi R^2 \bar{\rho}/3,
\label{Phi} 
\end{equation}
and where, to the approximation in which we are working (i.e.,
dropping terms of order $R^3$ and higher), we have $R^2 = X^2 + Y^2 +
Z^2$. Thus, $\Phi$ satisfies the ordinary Poisson equation
\begin{eqnarray} 
\nabla^2 \Phi & = & \nabla^2 \phi + 4 \pi \bar{\rho} \nonumber \\
& = & 4 \pi (\delta \rho + \bar{\rho}) \nonumber \\
& = & 4 \pi \rho.
\label{Poisson}
\end{eqnarray}
When $\Lambda = 0$, Eq.~(\ref{n2}) is precisely the usual form of
Newtonianly perturbed Minkowski spacetime (see, e.g., Sec.~4.4a of
\cite{wald}).  Thus, in the spacetime of Eq.~(\ref{metric}), when
$\Lambda = 0$, Newtonian gravity holds to a very good approximation in
the vicinity of any observer following the Hubble flow, where ``in the
vicinity'' here means on scales much smaller than the Hubble
radius. Even when $\Lambda \neq 0$, if $|\delta \rho| \gg \bar{\rho}$
in the neighborhood of the observer, realistic values of $\Lambda$
have $\Lambda R^2 \ll \Phi$ out to distances much smaller than the
Hubble radius. Thus, Newtonian gravity holds to an excellent
approximation in the vicinity of such observers as well.

In summary, we may characterize our cosmological model of
Eq.~(\ref{metric}), together with Eqs.~(\ref{d1phi})--(\ref{d3phi}),
as one which corresponds closely to a Robertson-Walker model as far as
the Hubble flow of the matter and the causal structure of the
spacetime are concerned, but in which the local distribution of matter
may be highly inhomogeneous.  In addition, as we have just noted, on
scales small compared with those set by the underlying
Robertson-Walker model, Newtonian gravity holds to a very good
approximation. Apart from negligibly small regions of spacetime which
contain black holes or other strong field objects, we believe that our
universe is accurately described by this model. In any case, our model
is a relatively precise, mathematically consistent cosmological model
which describes the spacetime structure and distribution of matter on
all scales, and is not in obvious conflict with any observed
properties of our universe.

\subsection{Propagation of Photon Beams}
\label{Propagation of Photon Beams}

Let us now consider this cosmological model from the perspective of
photons ($\equiv$ null geodesics) propagating in it, and compare this
to what photons would encounter in a Robertson-Walker model. All
gravitational focusing and shearing effects on an infinitesimal beam
of light rays in the vicinity of a null geodesic $\gamma$ are
described by the geodesic deviation equation (see, e.g. \cite{wald})
\begin{equation} 
\frac{d^2 \eta^a}{d \lambda^2} = - {R_{bcd}}^a k^b k^d \eta^c,
\label{gd}
\end{equation}
where $k^a$ is the tangent to $\gamma$ corresponding to affine
parameter $\lambda$, and $\eta^a$ is the deviation vector to an
infinitesimally nearby null geodesic in the beam. The Riemann
curvature tensor appearing in Eq.~(\ref{gd}) can be decomposed into its
Ricci and Weyl pieces in the usual way (see, e.g. \cite{wald})
\begin{equation} 
R_{abcd} = C_{abcd} +\left(g_{a[c}R_{d]b}-g_{b[c}R_{d]a}\right)
-{1\over3}Rg_{a[c}g_{d]b}.
\label{wr}
\end{equation}
The Ricci curvature directly produces a rate of change of convergence
of the beam of geodesics, while the Weyl curvature directly produces
a rate of change of shearing. 

In a Robertson-Walker model the Weyl tensor vanishes and, by
Einstein's equation, the Ricci tensor is of the form $R_{ab} = 8 \pi
(T_{ab} - 1/2\,T g_{ab})$, with $T_{ab}$ given by Eq.~(\ref{Tab}). The
geodesic deviation equation then takes the form 
\begin{equation} 
\frac{d^2 \eta^a}{d \lambda^2} = - 4 \pi \omega^2 \rho \eta^a,
\label{gd2}
\end{equation}
where $\omega$ is the frequency of the photon as measured in the
Robertson-Walker rest frame. This corresponds to a steady increase in
the convergence of the beam of geodesics, with no shear. Contrast
this behavior with the propagation of photons in the cosmological
model of Eq.~(\ref{metric}) in the case where the matter is highly
clumped on various scales, but with no (or negligible) matter
distributed between the clumps. In this case, the Ricci tensor
vanishes along the geodesic, except for rare instances when the photon
propagates through a clump of matter.  On these rare occasions, the
Ricci curvature briefly becomes extremely large compared with that of
the underlying Robertson-Walker model. The Weyl curvature also will be
small except in similarly rare instances of propagation through (or
very near) a sufficiently dense clump of matter. Thus, when the matter
distribution is highly clumped, at almost all times the propagation of
a beam of photons in the spacetime of Eq.~(\ref{metric}) would be
indistinguishable from propagation in flat spacetime. Occasionally,
however, the beam may receive a strong ``kick'' of Weyl and/or Ricci
curvature. Thus, the local history of a photon propagating in the
spacetime of Eq.~(\ref{metric}) could hardly be more different from
the local history of a photon propagating in a Robertson-Walker model!
Nevertheless, there are some global correspondences. In particular,
since the causal structure of the spacetime of Eq.~(\ref{metric})
corresponds closely to that of the underlying Robertson-Walker metric,
at each redshift\footnote{Since $|\phi|\ll1$ and the velocity of matter
relative to the Hubble flow is small, we neglect the difference
between redshifts in the metric of Eq.~(\ref{metric}) and in the
underlying Robertson-Walker model.}
the area of the boundary of the past of an event in
the spacetime of Eq.~(\ref{metric}) must be very nearly equal to the
area of the past light cone of the corresponding event in the
underlying Robertson-Walker metric. We will return to this point in
Subsection~\ref{ss:checks}.

In order to calculate magnification and shear effects on a (nearly)
point source due to gravitational lensing, we need to integrate the
geodesic deviation equation (Eq.~(\ref{gd})) along a null geodesic
connecting the source to the observer. To do this, we need to know the
curvature along the geodesic. The curvature is determined directly by
a knowledge of the underlying Robertson-Walker model together with
$\phi$. We will assume that, in the underlying Robertson-Walker model,
the distance scales set by the spatial curvature and $\Lambda$ are at
least as large as the Hubble radius, $R_H$. The spacetime curvature of
the Robertson-Walker model is then of order $1/R_H{}^2$.
Contributions of $\phi$ to the spacetime curvature which are smaller
than $1/R_H{}^2$ will therefore be neglected. From
Eq.~(\ref{poisson}), together with the assumption that $\phi$ is
bounded and has vanishing spatial average, it follows that $\phi$ is
uniquely determined by specifying the matter distribution $\delta
\rho$.  However, Eq.~(\ref{poisson}) is a nonlocal equation, so in
principle the locally encountered curvature could depend upon the
distribution of matter in arbitrarily distant parts of the
universe.\footnote{Note that since, for an open universe, $\delta
\rho$ does not fall off to zero at infinity, we cannot assume, a
priori, that $\phi$ is given in terms of $\delta \rho$ by the usual
Poisson integral expression that would hold for a localized mass
distribution.} Nevertheless, we shall now argue that, under our
cosmological assumptions, only the distribution of matter within $R_H$
is relevant.

\subsection{The Local Nature of the Influence of Matter on Photon Beams}
\label{The Local Nature}

Let $S$ be a sphere of proper radial distance $R_H$
centered about the point $x$ at which we wish to evaluate $\phi$. Let
$G_D(x,x')$ denote the Dirichlet Green's function for the equation
$a^{-2} h^{ab} D_a D_b G(x,x') = - 4 \pi \delta(x,x')$ for the region
enclosed by $S$. (A simple, explicit formula for $G_D$ in the case of
flat geometry can be found, e.g., in Sec.~2.6 of~\cite{jack}.)
Then, by Green's identity, we have
\begin{equation} 
\phi(x) = - \int_V G_D (x,x') \delta \rho (x') dV'
- \frac{1}{4 \pi} \int_S \phi (x') \hat{r}'{}^a D'_a G_D (x,x')\,dS',
\label{gi}
\end{equation}
where the volume integral extends only over the region enclosed by
$S$. Under our above assumptions, the contribution of $\phi$ to the
curvature is given directly in terms of the second spatial derivatives
of $\phi$, since the contributions from the time derivatives of
$\phi$, products of first derivatives of $\phi$, etc., have been
assumed to be negligible compared with the linear contributions from
the second spatial derivatives of $\phi$. Differentiating
Eq.~(\ref{gi}), we obtain
\begin{eqnarray} 
D_a D_b \phi (x) &=& - \int_V D_a D_b G_D (x,x') \delta \rho (x')\,dV' 
\nonumber \\ 
& & \mbox{} - \frac{1}{4 \pi}
\int_S \phi (x') \hat{r}'{}^a D'_a D_a D_b G_D (x,x')\,dS'.
\label{gi2}
\end{eqnarray}
However, the surface term in Eq.~(\ref{gi2}) is of order
$|\phi|/R_H{}^2$, and thus, in view of Eq.~(\ref{d1phi}), it can be
neglected. Therefore the curvature at $x$ is determined by the
matter distribution only within a Hubble radius of $x$, as we desired
to show. It should be emphasized that this conclusion is {\em not} a
consequence of any causality arguments but, rather, follows directly
from our above {\em assumption} that $\phi$ is small at distances of
order $R_H$, as is necessary for the underlying Robertson-Walker metric 
to be a good description of spacetime structure on cosmological scales. 

We now make the additional assumption that there is a (co-moving)
scale ${\cal R} \ll R_H$ such that no strong correlations in the
distribution of matter occur on scales greater than ${\cal R}$. Under
these circumstances it seems clear that the curvature at a given point
can be accurately calculated---at least for the purposes of
determining geodesic deviation---by taking into account only the
matter distribution within a distance ${\cal R}$ of that point. We
have not attempted to give a precise formulation or proof of this
claim, but a justification for it can be given as follows. First we
note that, by Einstein's equation, the Ricci curvature is determined
by the matter distribution in a completely local manner. Therefore,
matter can have a nonlocal influence on a photon beam only via Weyl
curvature.  To calculate the Weyl curvature associated with a
distribution of matter we need to evaluate the trace-free part of the
second derivatives of $\phi$, as given by Eq.~(\ref{gi2}) with the
surface term omitted. We break up the volume $V$ in Eq.~(\ref{gi2})
into a union of regions of size ${\cal R}$, excluding the ball of
radius ${\cal R}$ centered at $x$. In the case of flat spatial
geometry, each of these regions will make a contribution of order
$m/D^3$ to the Weyl tensor at $x$, where $D$ is the distance of the
region from $x$, and $m$ is of the order of the expected mass,
$\bar{\rho}\,{\cal R}^3$, contained in that region. However, by our
assumption, there will be no correlations between the contributions
from the different regions. Hence, by a simple ``random walk''
estimate, we find that the total contribution to the Weyl tensor at
$x$ from all of $V$ except for the ball of radius ${\cal R}$ centered
at $x$ should be no greater than $\sim m/{\cal R}^3 \sim
\bar{\rho}$. Similar estimates hold if the geometry is curved or a
cosmological constant is present, since $G_D$ will differ
significantly from the flat case only at distances comparable to
$R_H$, and the contributions from these regions should be negligible.

We note that $\bar{\rho}$ is the same order of magnitude as the
curvature of the underlying Robertson-Walker metric. A Ricci curvature
of this magnitude and having a consistent sign (as occurs in the
Robertson-Walker model) could have a significant effect on the
convergence of a beam of photons propagating over cosmological
distances. However, a randomly fluctuating Weyl curvature of this
magnitude should have a completely negligible effect upon the shear
(merely adding a tiny bit of ``noise'' to the Weyl curvature resulting
from nearby matter), and an even smaller effect upon the
convergence. Thus, no significant error should be made by considering
only the curvature resulting from the presence of matter within ${\cal
R}$ of the photon path, as we desired to show.

Since we have assumed that ${\cal R} \ll R_H$ and that the distance
scales set by the spatial curvature and/or $\Lambda$ are at least as
large as $R_H$, the Dirichlet Green's function within ${\cal R}$ of
$x$ will be well approximated by $1/r$, where $r$ denotes the proper
distance between $x$ and $x'$. Thus, Eq.~(\ref{gi})---with the
surface term omitted and the volume integral restricted to a ball of
radius ${\cal R}$ around $x$---reduces to the usual Poisson integral
formula, and the curvature can be obtained from formulas arising from
ordinary Newtonian gravity (see Sec.~\ref{method} below). It is
somewhat more convenient to work with the potential $\Phi$ of
Eq.~(\ref{Phi}) rather than $\phi$. It follows that $\Phi$ is given by
the usual Poisson integral formula of $\rho$ (rather than $\delta
\rho$) over the region enclosed by ${\cal R}$.

\subsection{Our Method}
\label{Our Method}

The basic idea of our procedure in its most general context can now be
explained. We choose an underlying Robertson-Walker model and
(co-moving) scale, ${\cal R}$, with ${\cal R} \ll R_H$ in the present
universe.\footnote{More generally, we could specify a probability
distribution for ${\cal R}$, although we shall not do so in this
paper.} We then specify a probability distribution for how the matter
is distributed within ${\cal R}$. This probability distribution may
vary with cosmological time; it is constrained only by the requirement
that the average amount of mass contained within ${\cal R}$ agree with
that occurring in the underlying Robertson-Walker model.  We then
perform a ``Monte Carlo'' propagation of a beam of photons backward in
time, starting from the present, in the following manner: We prescribe
a matter distribution (chosen from our probability distribution) in a
ball of radius ${\cal R}$. We calculate the Newtonian potential for
this matter distribution, and the corresponding curvature. Then we
choose a random impact parameter for the entry of a photon into this
ball, and we integrate Eq.~(\ref{gd}) through the ball. (In this step,
we take the photon trajectory to be a ``straight line'', i.e., we do
not attempt to include the (completely negligible) corrections due to
the tiny bending angle.) When the photon exits from this ball, we use
the underlying Robertson-Walker model to update the frequency of the
photon relative to the local rest frame of the matter, and to update
the proper radius corresponding to the comoving scale ${\cal R}$. Then
we choose a matter distribution in a new ball of comoving radius
${\cal R}$, choose another random impact parameter for entry of the
photon into this ball,\footnote{Note that, in general, this would
require the balls to overlap slightly. We neglect this overlap in our
analysis.}
and repeat the above calculations. We continue until the
photon has reached the desired redshift. By repeating this sequence of
calculations a large number of times---for most of our models we
performed about 2,000 such ``runs''---we build up good statistics on
what happens to beams of photons on our past light cone. From this we
obtain, for any given model, good statistical information on the
magnification, shear, and rotation of images of (nearly) point sources
at any redshift. We will spell out the details of our procedure more
explicitly in the next section.

In comparison with other approaches, ours most closely resembles the
``Swiss cheese'' models, wherein one takes a matter dominated
Robertson-Walker model, removes the dust from spherical balls, and
redistributes the mass within these balls in some other (arbitrarily
chosen) spherically symmetric manner. However, it differs from the
Swiss cheese models in the following significant ways: (i) The
``cheese'' has been completely eliminated. (ii) The mass within a
given ball need not be equal to the corresponding Robertson-Walker
mass, though equality must still hold on average. (iii) The matter
distribution within the balls need not be spherically symmetric. (iv)
We do not consider the propagation of photons in a single, fixed
cosmological model. Rather, each photon in effect ``creates its own
cosmological model'' via our Monte Carlo procedure during the course
of its propagation. (v) Although it is not a necessary facet of the
Swiss cheese models, most analyses of the Swiss cheese
models~\cite{kantowski,dyerroeder1} have attempted to calculate only
averages of certain lensing quantities, and, in the course of doing
so, have imposed an ``opaque radius'' cutoff---within which photons
are absorbed---which biases the
results towards defocusing relative to Robertson-Walker models. Our
analysis determines the probability distributions for magnification,
shear, and rotation of sources by doing an exact, Monte Carlo
calculation, imposing no opaque radius cutoff. As we shall see, our
results show no bias towards defocusing relative to the underlying
Robertson-Walker model, provided that all of the high luminosity
images are included (see Sec.~\ref{supernovae} for further discussion).

Our approach also bears some similarity to analyses which start with a
model of the matter distribution in the universe---obtained
analytically~\cite{schneider,SEF,rauch} or from N-body
codes~\cite{matzner,wambsganss2,wambsganss,tomita}---and
then project the matter
into lens planes lying between the source and observer. Ray shooting
methods are used to numerically obtain bending angles of a large sample of
photons, from which the amplification and shear distribution of images
can then be computed. Our approach uses the geodesic deviation
equation rather than the lens equation and is considerably simpler and
more flexible. It also avoids any artifacts resulting from putting all
the matter into lens planes.

\subsection{The Relevant Scales of Clustering and Clumping}
\label{ss:relevant_scales}

Two final issues remain to be addressed: (1) What clustering scale
${\cal R}$ should be chosen to adequately model statistical lensing
effects in our universe, i.e., what is the largest scale on which the
clustering of matter has an important effect upon lensing? (2) On what
scales (below ${\cal R}$) does one have to model the details of the
matter distribution in order to adequately treat statistical lensing
effects, i.e., what is the smallest scale on which the clumping of
matter has an important effect?

In analyzing these questions, it is convenient to view galaxies as the
basic ``building blocks'' of the distribution of matter in the
universe. (Although we do not exclude the possibility that
substantial amounts of matter may be distributed between galaxies,
we assume that such matter is distributed in a
relatively uniform way.) It is essential to take into account
the clumping of matter on the scale of galaxies in order to adequately
model lensing effects. In essence, the first question above asks to
what extent the clustering of the galaxies themselves must be taken
into account, while the second question asks to what extent the clumping
substructure of the matter within galaxies must be taken into account.

As already noted, it follows from Einstein's equation that the Ricci
curvature is determined by the matter distribution in a completely
local manner. The effects of Ricci curvature on lensing should
therefore depend only upon the density contrasts associated with
galaxies, and not upon the ``shape'' of galaxies. This will be
verified explicitly in Subsection~\ref{ss:checks}. Furthermore,
these Ricci curvature effects should depend only weakly on the
clustering of galaxies, since the clustering should merely produce
some correlations in the times of passage of a photon through
different galaxies, and these effects should largely ``wash out'' over
cosmological distance scales. Thus, we believe that the clustering of
galaxies should have a negligible influence on lensing effects
produced by Ricci curvature.

On the other hand, simple estimates show that the Weyl curvature of
a spherical aggregate of matter of mass $m$ and radius $r$ can have a
substantial effect on lensing only if the matter ``lies within its own
Einstein radius'', $r_E$, i.e., only if
\begin{equation} 
r^2 \lap r_E{}^2 \sim m D,
\label{er}
\end{equation}
where $D$ denotes a cosmological distance and we use units where $G =
c = 1$. Note that this relationship is marginally satisfied by
individual galaxies (or at least by their central cores), so the
Weyl curvature of individual galaxies can (at least occasionally)
produce significant lensing effects. Clustering of galaxies can
produce important Weyl curvature effects only in
circumstances when the clusters themselves
satisfy Eq.~(\ref{er}). This {\em does} occur in the central portions
of rich clusters of galaxies, so the effects of clustering cannot
always be assumed to be negligible.

However, in the limit where galaxies can be treated as ``point
masses''---as occurs if Eq.~(\ref{er}) is satisfied by a wide
margin---it follows from the analysis given in
Subsection~\ref{ss:clustering} below that even very strong clustering
of the galaxies will have at most a tiny effect on the lensing
probability distributions for the magnification, shear, and rotation
of (nearly) point sources. (On the other hand, clustering {\em would}
still have an important effect on some lensing quantities, such as
bending angles, which we do not calculate here.) Thus, clustering
effects can be of importance for the statistical lensing quantities
treated here only only when individual galaxies fail to satisfy
Eq.~(\ref{er}), but these galaxies form clusters which satisfy
Eq.~(\ref{er}) (at least in their core regions). In these
circumstances the neglect of the clustering of galaxies should
underestimate the lensing effects somewhat. However, we do not believe
that such circumstances arise frequently enough to have an important
influence on the statistical lensing quantities we calculate.
Furthermore, as we shall conjecture in Subsection~\ref{ss:conjecture},
the point mass results should provide a firm upper limit to lensing
effects, even when galactic clustering is present.

Consequently, in this paper we shall take ${\cal R}$ to be the scale
of the separation between galaxies, thereby neglecting lensing effects
resulting from the clustering of galaxies. For the reasons detailed
above, we do not expect that this will result in any significant
errors in our calculations of the probability distributions for
magnification, shear, and rotation of images of cosmologically distant
sources. Some evidence in favor of this expectation will be given in
Section~\ref{supernovae}, where we will obtain results in close
agreement with~\cite{wambsganss}, despite our neglect of the effects
of clustering.\footnote{This expectation could be further tested by
re-doing our analysis taking ${\cal R}$ to be the scale of separation
of clusters of galaxies and using appropriately chosen probability
distributions for the distribution of mass within clusters. We have
not yet attempted to carry out such an analysis.}

We turn now to the issue of how small a scale of clumping of matter we
must consider in order to calculate gravitational lensing effects. In
principle the clumping of matter on arbitrarily small scales
(including atomic and sub-atomic scales) could have an important
effect on lensing---though we would have to use physical, rather than
geometric, optics to calculate these effects when the objects are so
small that the scale of variation of the gravitational field becomes
less than the wavelength of the light. However, the finite size of the
source which is being lensed provides an effective cutoff to lensing
produced by clumping on small scales. This follows because the lens
merely magnifies (as well as shears and rotates) the image of the
source, keeping the surface brightness constant~\cite{SEF}. Thus, if
the angular size of the (assumed to be uniform) source is much larger
than the angular scale associated with the lens, the lensing effects
caused by clumping should have little effect, as only a relatively
small part of the source would be magnified by the presence of a clump
of matter (and the rest of the source may be correspondingly
demagnified by the absence of matter between clumps). In other words,
the net angular size of the image of a source of finite size will not
be significantly affected by sufficiently small scale lensing, and
consequently, the luminosity of the image also will not be greatly
affected.\footnote{If the angular size of the source is much larger
than the angular scale of separation between the clumps of matter,
then the lensing effects of the matter should wash out completely.}
The angular scale of the source is $\sim r_S/D_S$, where $r_S$ denotes
the size of the source and $D_S$ denotes its distance, and the angular
scale associated with the lens is $\sim r_E/D_L$, where $D_L$ denotes
the distance of the lens. Taking $D_S$ and $D_L$ to be cosmological in
scale and using Eq.~(\ref{er}) for $r_E$, we find that lensing effects
should not be important unless the mass of the lens satisfies
\begin{equation} 
m \gap r_S{}^2 /D.
\label{mlim}
\end{equation}
The smallest sources of interest here (central regions of quasars and
supernova shells at an early stage of expansion) have $r_S \gap
10^{-3}$ light years, so taking $D \sim 10^{10}$ light years, we find
\begin{equation} 
m \gap 10^{-3} M_\odot.
\label{mlim2}
\end{equation}
Clumping of matter on mass scales smaller than
Eq.~(\ref{mlim2}) should not be relevant for the sources we
consider. However, the clumping of matter down to the scale
of Eq.~(\ref{mlim2}) is potentially of importance. In particular, the
clumping of matter in galaxies into stars can have a significant effect
upon the probability distribution for the magnification of light
emitted from quasars and supernovae.

Fortunately, it is not necessary to model a galaxy as $10^{11}$ or so
point mass stars in order to calculate its lensing effects. The
clumpiness of matter will be relevant only very close to the path of
the photon. If, say, we let $d \sim 100 r_E$, where $r_E$ denotes the
Einstein radius of a single star, then the discreteness of the
galactic mass distribution due to stars which lie outside of a tube of
radius $d$ around the photon path can be ignored, i.e., outside of the
tube the galactic matter distribution can be treated as
continuous. Consequently, in our analysis we will take account of all
``microlensing'' effects due to small scale clumping of matter (say,
into stars) in the following manner: First, we model the galaxy as a
continuous mass distribution and compute its Newtonian
potential. Then, when a photon passes through the galaxy in our Monte
Carlo simulations, we remove the continuous galactic matter lying
within cylindrical radius $d$ of the path (or the portion of this
matter assumed to be clumped into stars), and subtract the Newtonian
potential of this removed matter. Finally, we randomly redistribute
this removed mass back into the cylinder in the form of stars, and we
add in the Newtonian potential of these ``point masses''. In this
manner we take full account of the small scale clumping of matter in a
computationally efficient way.

\section{Details of Our Method}
\label{method}
In the previous section we spelled out our cosmological assumptions
and described our method for calculating statistical lensing effects
on cosmological sources. The purpose of this section is to provide
a more concrete and explicit description of our approach.

As discussed in the previous section, we first must choose an
underlying matter dominated Robertson-Walker model, which we may
characterize by the parameters ($H_0$,$\Omega_0$,$\Lambda$), where the
subscript ``$0$'' denotes the present value, and where $\Omega_0
\equiv 8 \pi \bar{\rho}_0/3 H_0$. (The value of $k$ can then
determined from Einstein's equation (\ref{ee2'}).)  For the
calculations in this paper, we use the value $H_0 = 70\
\mbox{km}\,\mbox{s}^{-1}\,\mbox{Mpc}^{-1}$. We choose a comoving
scale, ${\cal R}$, which, as discussed above, we take to be the scale
of the galactic separations. For most of the calculations in this
paper the value of ${\cal R}$ will be taken to correspond to $2\
\mbox{Mpc}$ in the present universe.  Next we specify how mass is
distributed in (and in between) galaxies. As explained in the previous
section, in general, we may specify this as a probability distribution
which is subject only to the constraint that, on average, the mass
assigned to the galaxy equals the mass contained in a ball of radius
${\cal R}$ in the underlying Robertson-Walker model. In the present
paper, however, we will only be concerned with simple ``toy models''
for the mass distribution, and, in each model universe, we will take
all galaxies to have identical mass and structure, rather than
specifying a probability distribution. (However, we will consider
model universes with a wide range of different galactic mass
distributions.) We also will choose the galactic mass distributions
(expressed in terms of proper---not comoving---distances) not
to vary with cosmological time.

Our basic strategy for determining lensing effects can be explained in
the following manner. Imagine a telescope at an event $p$ in the
present universe which is pointed in some direction in the sky, so
that it only accepts photons which impinge upon it with null tangent
very close to the direction $k^a$. Suppose, further, that $k^a$ has
been chosen so that, when followed backwards in time, a photon
arriving at event $p$ with null tangent $k^a$ would have emerged from
a source at redshift $z$, i.e., suppose that the telescope happens to
be ``pointed at'' a source at redshift $z$. If the source is
sufficiently small, the photons emitted by it which are accepted by
the telescope can be treated as a beam of null geodesics which all lie
on the past light cone of $p$ and have directions differing only
infinitesimally from $k^a$. The deviation vector characterizing any
given geodesic in this beam must therefore vanish at $p$, and is
uniquely determined by its time derivative at $p$. The relationship
between the image produced by the telescope and the actual structure
of the source is determined by the relationship between the time
derivatives of the deviation vectors in this infinitesimal beam at $p$
and the values of these deviation vectors at redshift $z$. To
calculate this relationship we must integrate the geodesic deviation
equation (\ref{gd})---{\em backwards} in time starting from $p$---for
the 2-dimensional space of spatial deviation vectors $\eta^a$ which
are orthogonal to $k^a$ and which vanish at $p$. By linearity of the
geodesic deviation equation, the components of these deviation vectors
in a parallely propagated frame are determined by a $2 \times 2$
matrix, ${\cal A}^\mu{}_\nu(\lambda)$, defined by
\begin{equation} 
\eta^\mu (\lambda) =
{\cal A}^\mu{}_\nu(\lambda) \frac{d \eta^\nu}{d \lambda}(0),
\label{Amunu}
\end{equation}
where $\lambda = 0$ corresponds to the present time. It follows
immediately from the geodesic deviation equation~(\ref{gd}) that
${\cal A}^\mu{}_\nu(\lambda)$ satisfies
\begin{equation}
{d^2{\cal A}^\mu{}_\nu\over d\lambda^2}=
-R_{\alpha\beta\sigma}{}^\nu k^\alpha k^\sigma {\cal A}^\beta{}_\mu,
\label{maineq}
\end{equation}
with initial conditions 
${\cal A}^\mu{}_\nu(0) = 0$ and $d{\cal A}^\mu{}_\nu(0)/d \lambda =
\delta^\mu{}_\nu$. Equation (\ref{maineq}), together with these initial
conditions, uniquely determines ${\cal A}^\mu{}_\nu$ for all $\lambda$. 

All lensing effects considered in this paper are obtained directly
from ${\cal A}^\mu{}_\nu$ as follows.  Let $\lambda_z$ denote the
affine parameter of the beam of photons at redshift $z$. We decompose
the matrix ${\cal A}^\mu{}_\nu(\lambda_z)$ as a product of a proper
(i.e., unit determinant) orthogonal matrix $O^\mu{}_\nu$ and a
self-adjoint matrix $S^\mu{}_\nu$
\begin{equation} 
{\cal A}^\mu{}_\nu(\lambda_z) = O^\mu{}_\sigma S^\sigma{}_\nu.
\label{OS}
\end{equation}
The matrix $S$ can be characterized by its two eigenvalues, $\alpha_1$
and $\alpha_2$. The area, $A$, of the beam at redshift $z$ is given by
\begin{equation} 
A = \alpha_1 \alpha_2 = \det {\cal A}.
\label{area}
\end{equation}
The corresponding magnification---and, hence, amplification---of the
(nearly point) source at redshift $z$ (relative to a source placed at
the same affine parameter in flat spacetime) is given by
\begin{equation} 
\mu = \lambda^2_z/A.
\label{mag}
\end{equation}
Note that a caustic in the beam of geodesics occurs precisely when
$\det{\cal A} = 0$, in which case the magnification/amplification 
of a point source becomes infinite. 

The distortion of the image is characterized by the ratio, $\epsilon$,
of the eigenvalues of ${\cal A}$
\begin{equation} 
\epsilon = |\alpha_1/\alpha_2|,
\label{epsilon}
\end{equation}
where we take $|\alpha_1| \leq |\alpha_2|$, so that $0 \leq \epsilon
\leq 1$. For a circular source, the image will be elliptical in
shape, with axial ratio given by $\epsilon$. This distortion also can
be characterized by the net shear\footnote{The term ``shear'' is also
commonly used for the optical tensor $\sigma_{ab}$ (see, e.g.,
(9.2.28) of \cite{wald}) and/or its magnitude, $\sigma$. These
quantities characterize the ``rate of shearing'' of the beam, and
should be clearly distinguished from the ``net shear'' defined here.}
of the image, defined by
\begin{equation} 
\gamma = \frac{|\alpha_1 - \alpha_2|}{2\lambda_z}.
\label{shear}
\end{equation}

If $A > 0$, corresponding to a beam which has not passed through a
caustic (or has passed through an even number of caustics), we may
choose $S$ to be positive definite (as opposed to negative
definite). This uniquely fixes $O$. The matrix $O$ can be
characterized, in turn, by a rotation angle $\Theta$ ranging between
$- \pi$ and $\pi$, which may be interpreted as the angle of rotation
of the image relative to the orientation that the image would have
had in the underlying
Robertson-Walker model. Such a rotation results from the cumulative effects of
shearing in different directions produced by masses lying in different
``lens planes''; lensing by masses lying in a single plane orthogonal
to the photon trajectory would not yield any rotation. If $A < 0$,
there is a sign ambiguity in $S$, and a corresponding sign ambiguity
in $O$. We resolve this ambiguity (arbitrarily) by choosing $\Theta$
to range between $- \pi/2$ and $\pi/2$

In this paper, the desired statistical information on the
magnification, shear, and rotation of sources at redshift $z$ will be
obtained by repeated ``Monte Carlo'' integration of
Eq.~(\ref{maineq}), as explained in detail below. It is worth noting
that Eq.~(\ref{maineq}) corresponds to the ``primitive form'',
(Eq.~(\ref{gd})), of the geodesic deviation equation, rather than the
mathematically equivalent ``optical equations'' form (see, e.g.,
Eqs.~(9.2.32)--(9.2.33) of \cite{wald}) used in many other
analyses. For our purposes, Eq.~(\ref{maineq}) has a
significant advantage over the optical equations in that ${\cal
A}^\mu{}_\nu$ varies continuously when caustics occur, whereas
quantities---such as the convergence of the congruence---appearing
in the optical equations become singular at caustics.

We begin our backwards evolution of Eq.~(\ref{maineq}), with initial
conditions ${\cal A}^\mu{}_\nu(0) = 0$ and $d{\cal A}^\mu{}_\nu(0)/d \lambda =
\delta^\mu{}_\nu$, by imagining that the beam of photons enters a ball
of radius ${\cal R}$ with a randomly chosen impact parameter, $b$,
i.e., we take $b = \sqrt{p}\,{\cal R}$ with $p$ chosen randomly
from the interval $[0,1]$. As described above, the mass distribution
in this ball has been prescribed, and we can calculate the Newtonian
potential, $\Phi$, corresponding to this mass distribution by solving
the ordinary Poisson equation (see Eq.~(\ref{Poisson}) above). The
relevant components of the Riemann curvature tensor can be calculated
straightforwardly for the metric of Eq.~(\ref{n2}), yielding
\begin{eqnarray}
R_{abcd}k^bk^d & = &  2 \omega^2 \partial_a\partial_c\Phi
+k^b\partial_b(k^d\partial_d\Phi)\eta_{ac} \nonumber \\
& = & \omega^2  [2 \partial_a\partial_c\Phi
+Z^b\partial_b(Z^d\partial_d\Phi)\eta_{ac}].
\label{riem}
\end{eqnarray}
Here $\eta_{ab}$ denotes the ordinary Minkowski metric associated with
the coordinates $(T,X,Y,Z)$ (see Eq.~(\ref{n2}) above), $\partial_a$
denotes the derivative operator of $\eta_{ab}$, $Z^a$ denotes a unit
vector in the ``$Z$-direction'' ($\equiv$ the direction of propagation
of the photon), and $\omega = dT/d\lambda$ denotes the frequency of
the photon. Without loss of generality, we choose our affine
parametrization so that initially $\omega = 1$. Note that the
$\Lambda$ term in Eq.~(\ref{n2}) does not contribute to the relevant
components of the Riemann tensor, i.e., we have {\em not} assumed in this
equation that $\Lambda = 0$.

In integrating Eq.~(\ref{maineq}) through the ball, we neglect the tiny
bending angle of the photon trajectory, and we also neglect the tiny
changes in ${\cal A}^\mu{}_\nu$ occurring as the photon traverses the
ball. Thus, when the beam of photons exits from the ball, the new
values of ${\cal A}^\mu{}_\nu$ and $d{\cal A}^\mu{}_\nu(0)/d \lambda$ are related
to the values they had when entering the ball as follows:
\begin{eqnarray}
&({\cal A}^\mu{}_\nu)_1 = ({\cal A}^\mu{}_\nu)_0 + \omega^{-1} \Delta Z (d{\cal A}^\mu{}_\nu /d
\lambda)_0&
\label{A1}  \\
&(d{\cal A}^\mu{}_\nu /d \lambda)_1 = (d{\cal A}^\mu{}_\nu /d \lambda)_0 -
\omega J^\mu{}_\beta ({\cal A}^\beta{}_\nu)_0,&
\label{A2}
\end{eqnarray}
where we have used the fact that $dZ/d\lambda = \omega$, and where we have
defined
\begin{equation}
J^\mu{}_\nu \equiv \int dZ\, [2 \partial^\mu \partial_\nu \Phi
+Z^b\partial_b(Z^d\partial_d\Phi)\delta^\mu{}_\nu],
\label{J}
\end{equation}
with the integral taken over the ``straight line'' photon path through
the ball. Since $({\cal A}^\beta{}_\nu)_0 = 0$, the $J^\mu{}_\nu$ term will
not contribute to Eq.~(\ref{A2}) for the traversal through
the first ball, but it will contribute for all subsequent balls.

For the cases considered in this paper, $J^\mu{}_\nu$ can be
calculated analytically. Specifically, if we align our axes so that the
photon propagates in the $Z$-direction and lies in the $Y = 0$ plane,
then for a ``point mass'' of mass $M$ placed at the center of the ball
(i.e., linearized Schwarzschild geometry), we have
\begin{eqnarray}
(J^P)^{X}{}_{X} &=& -2M\left(1-{b^2\over R^2}\right)^{1/2}
  \left({2\over b^2}+{1\over R^2}\right) \nonumber \\
(J^P)^{Y}{}_{Y} &=& +2M\left(1-{b^2\over R^2}\right)^{1/2}
  \left({2\over b^2}+{1\over R^2}\right) 
\label{JP} \\
(J^P)^{X}{}_{Y} &=& 0, \nonumber
\end{eqnarray}
where $b$ is the impact parameter and $R$ is the proper radius of the
ball ($ = {\cal R}$ for the initial ball).  For a uniform density ball
of mass $M$ and radius $d$ concentric with the ball of radius
$R$, we have for $b \leq d$,
\begin{eqnarray}
(J^U)^X{}_X &=& (J^P)^{X}{}_{X} + 4M\left(1-{b^2\over d^2}\right)^{1/2}
\left({1\over b^2}+{2\over d^2}\right) \nonumber \\
(J^U)^Y{}_Y &=& (J^P)^{Y}{}_{Y} - 4M\left(1-{b^2\over d^2}\right)^{1/2}
\left({1\over b^2}-{1\over d^2}\right) \label{JU} \\
(J^U)^X{}_Y &=& 0, \nonumber
\end{eqnarray}
whereas $(J^U)^\mu{}_\nu = (J^P)^\mu{}_\nu$ when $b > d$. Similarly,
for a truncated isothermal ball (with density given by $\rho\propto1/r^2$) of
mass $M$ and ``cutoff radius'' $d$, we have for $b \leq d$,
\begin{eqnarray}
(J^I)^X{}_X &=& (J^P)^{X}{}_{X}
+{4M\over b^2}\left(1-{b^2\over d^2}\right)^{1/2} \nonumber \\
(J^I)^Y{}_Y &=& (J^P)^{Y}{}_{Y}
-{4M\over b^2}\left(1-{b^2\over d^2}\right)^{1/2}
+ {4M\over bd}\cos^{-1}(b/d) \label{JI} \\
(J^I)^X{}_Y &=& 0, \nonumber
\end{eqnarray}
and, again, the point mass expressions apply when $b > d$.  Finally,
for a uniform density cylinder of cylindrical radius $d$ and
length $2R$, whose axis
passes through the center of the ball of radius $R$, when the photon
trajectory does not pass through the cylinder, we have
\begin{eqnarray}
(J^C)^X{}_X &=& - {2(1+\cos^2\alpha)\gamma\over\sin\alpha}
\left[{Z_+\over X_0{}^2+Z_+{}^2}-{Z_-\over X_0{}^2+Z_-{}^2}\right]
\nonumber \\ 
(J^C)^Y{}_Y &=& +
{2(1+\cos^2\alpha)\gamma\over\sin\alpha} \left[{Z_+\over
x_0{}^2+Z_+{}^2}-{Z_-\over X_0{}^2+Z_-{}^2}\right] \label{JC} \\
(J^C)^X{}_Y &=& 4\gamma X_0\cot\alpha \left[{1\over X_0{}^2+Z_+{}^2}-
{1\over X_0{}^2+Z_-{}^2}\right], \nonumber
\end{eqnarray}
where we have neglected the tiny ``end effects'' resulting from
the finite length of the cylinder. In Eq.~(\ref{JC}), the axes have
been aligned so that the photon again propagates in the $Z$-direction,
but now the $X$ and $Y$ axes are chosen so that the shortest line
connecting the photon path to the axis of the cylinder points in the
$X$-direction. The length of this shortest line is $X_0$, whereas
$\alpha$ denotes the angle between the path of the photon and the axis
of the cylinder. The quantities $Z_-$ and $Z_+$ denote, respectively, the
$Z$-values of the entry and exit of the photon from the ball of radius
$R$, with $Z=0$ being the point of closest approach of the photon path
to the axis of the cylinder. Finally, $\gamma \equiv M/2R$ denotes the
mass per unit length of the cylinder. When the photon trajectory
passes through the cylinder, the additional term $16 \gamma
\left(d^2-X_0{}^2\right)^{1/2}/{(d^2\sin\alpha)}$ must be added to
$(J^C)^X{}_X$, but the other components of $(J^C)^\mu{}_\nu$ are not
changed.

The $x$ and $y$ axes of our parallely propagated frame will be rotated
by an angle $\theta$ with respect to the $X$ and $Y$ axes of the above
formulas. Under a rotation, $U$, by angle $\theta$, components of $J$
transform as
\begin{equation} 
J \rightarrow UJU^{-1}
\label{UJ}
\end{equation}
so that, explicitly, the components of $J$ in our parallely propagated
frame are given by
\begin{equation} 
J^x{}_x = \cos^2 \theta J^X{}_X + \sin^2 \theta J^Y{}_Y
- 2 \sin \theta \cos \theta J^X{}_Y,
\label{Jxx}
\end{equation}
and so on.

We return, now, to following the evolution (into the past) of our beam
of photons. After the beam exits from the first ball, it is assumed to
immediately enter a second ball. However, on account of the Hubble
expansion---or, rather, the Hubble contraction, since we are evolving
backwards in time---the proper radius, $R$, of the second ball will
now be smaller than ${\cal R}$ by the factor $a_1/a_0$, where $a_0$
and $a_1$ denote, respectively, the values of the scale factor of the
underlying Robertson-Walker metric at the times of entry into, and
exit from, the first ball. Similarly, the frequency of the photons
will be blueshifted by this same factor. After updating the values of
$R$ and $\omega$ in this manner, we send the beam of photons into the
second ball with a randomly chosen impact parameter, and with a
randomly chosen value of the orientation angle, $\theta$, with respect
to the parallely propagated frame (see Eq.~(\ref{Jxx})). (In the
cylindrical case, the orientation of the axis of the cylinder also is
suitably randomized.) When it exits the second ball, the values of
${\cal A}^\mu{}_\nu$ and $d{\cal A}^\mu{}_\nu /d \lambda$ are changed
via Eqs.~(\ref{A1}) and~(\ref{A2}). The values of $R$ and $\omega$ are
then also updated, and the beam of photons is subsequently taken to
enter a third ball. This procedure is repeated until the desired
redshift is reached. At each redshift the area, shear, and rotation of
the beam can be computed in the manner explained below Eq.~(\ref{OS}).

\begin{figure}
\begin{center}
\begin{picture}(0,0)%
\epsfig{file=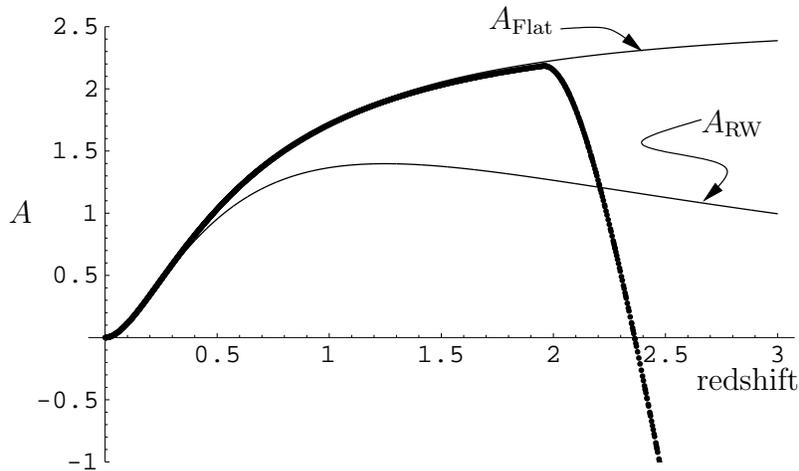}%
\end{picture}%
\setlength{\unitlength}{3947sp}%
\begingroup\makeatletter\ifx\SetFigFont\undefined%
\gdef\SetFigFont#1#2#3#4#5{%
  \reset@font\fontsize{#1}{#2pt}%
  \fontfamily{#3}\fontseries{#4}\fontshape{#5}%
  \selectfont}%
\fi\endgroup%
\begin{picture}(4947,2981)(4051,-5091)
\put(8401,-2911){\makebox(0,0)[lb]{\smash{\SetFigFont{12}{14.4}{\familydefault}{\mddefault}{\updefault}
$A_{\rm RW}$
}}}
\put(7066,-2266){\makebox(0,0)[lb]{\smash{\SetFigFont{12}{14.4}{\familydefault}{\mddefault}{\updefault}
$A_{\rm Flat}$
}}}
\put(4051,-3481){\makebox(0,0)[lb]{\smash{\SetFigFont{12}{14.4}{\familydefault}{\mddefault}{\updefault}
$A$
}}}
\put(8371,-4531){\makebox(0,0)[lb]{\smash{\SetFigFont{12}{14.4}{\familydefault}{\mddefault}{\updefault}
redshift
}}}
\end{picture}
\end{center}
\caption{The area vs. redshift of a beam of photons in a typical
Monte-Carlo run, for a universe with $\Omega_0 = 1$ and $\Lambda = 0$
in which all of the matter is in the form of point masses.}
\label{F:singlerun}
\end{figure}

Fig.~\ref{F:singlerun} shows the area of a beam of geodesics as a
function of $z$ for a typical ``run'', in the case of a universe with
$\Omega = 1$ and $\Lambda = 0$, and where the galaxies are treated as
point masses. The corresponding areas for the underlying
Robertson-Walker model and for a universe which is empty near the beam
(i.e, $R_{abcd} = 0$ in a neighborhood of the beam) are also shown
(see~\cite{dyerroeder2}). In this particular run, it can be seen that
the beam of photons propagates very nearly as though it were in empty,
flat spacetime, until it reaches a redshift $z \simeq 2$, at which
time it receives a strong ``kick'' due to a close encounter with a
(point mass) galaxy. A caustic then occurs at a redshift $z \simeq
2.4$.  To build up good statistics, we typically performed 2,000 such
``runs'' for each choice of cosmological model (i.e., values of
$\Omega_0$ and $\Lambda$) and galactic mass distribution. Our results
will be presented in the following sections.

Finally, we explain in more detail how the effects due to clumping of
matter into stars (or other sub-galactic structures) are
calculated. First, a continuous mass distribution is specified for
the galaxy, and the $J^\mu{}_\nu$ appropriate to this continuous
distribution was obtained (see Eqs.~(\ref{JP})--(\ref{JC}) above). A
cylindrical radius $d$ is then chosen so that a typical cylindrical
tube cutting through the galaxy contains at least $\sim 10$
``stars''. For the cases we consider, such a $d$ is automatically
much smaller than galactic scales, but much larger than the Einstein
radius of the individual stars. In our ``runs'', when a beam of
photons passes through a galaxy, we replace the $J^\mu{}_\nu$
calculated for the continuous mass distribution by ${J'}^\mu{}_\nu$
where
\begin{equation}
{J'}^\mu{}_\nu = J^\mu{}_\nu - 4\pi\sigma\,\delta^\mu{}_\nu
+ \sum_i ({J^P_i})^\mu{}_\nu (b_i, \theta_i).
\label{J'}
\end{equation}
Here $\sigma$ is the projected surface density (mass per unit area) of
the galaxy at the photon trajectory, $m_s$ denotes the mass of an
individual star, the ``impact parameters'' $b_i$ are chosen randomly
within a disk of radius $d$, and the rotation angles $\theta_i$ are
chosen randomly in $[0, 2 \pi]$. The sum ranges up to $N \approx
m/m_s$, where $m = \pi d^2 \sigma$ is the mass in a cylindrical tube
of radius $d$ about the photon path. The term
$-4\pi\sigma\,\delta^\mu{}_\nu$ subtracts the contribution to
$J^\mu{}_\nu$ from the continuously distributed mass within the
cylindrical tube, whereas the last term corresponds to adding back in
the contribution of this mass in the form of (point mass) stars, where
each $J^P_i$ is given by Eq.~(\ref{JP}) as modified by
Eq.~(\ref{Jxx}). (As we shall see in the next section, no essential
change in the probability distribution for ${J'}^\mu{}_\nu$ would
occur if we put the entire mass $m$ into a single ``star'' which is
randomly distributed in the disk of radius $d$.) We believe that all
statistical ``microlensing'' effects due to stars or other
sub-galactic structures can be accurately taken into account by this
procedure.

\section{Point Masses and Spherical Clustering}
\label{points_and_clustering}
In this section we will analyze lensing effects in the limiting case
where all of the matter in the universe is clumped into structures
which are much smaller than their own Einstein radii, so that these
objects may be treated as point masses. By a combination of analytic
and numerical arguments, we will establish---or, at least, present
strong evidence for---the following two key claims: (1) If the point
masses are randomly distributed throughout the universe, then the
probability distributions for magnification, shear, and rotation of
images depend, in an essential way, only on the total mass density of
the point masses. (This mass density, of course, is constrained to
equal the mass density of the underlying Robertson-Walker model.) In
other words, these probability distributions are (virtually)
independent of the masses of the individual point masses; in
particular, they are (virtually) the same for a universe randomly
populated by stellar mass black holes (or stars) as for a universe
randomly populated by galactic mass black holes. (Some partial results
along this line are given on P. 329 of \cite{SEF}.) (2) If the point
masses are not randomly distributed but are clustered on scales $\ll
R_H$ in a spherical but otherwise arbitrary manner, then the
probability distributions for magnification, shear, and rotation of
images are very nearly the same as for the randomly distributed case.

Taken together, results (1) and (2) provide a great
strengthening of the familiar claim that the ``optical depth'' for a
strong gravitational lensing event produced by point masses depends
only on the total mass density of these point
masses~\cite{pressgunn}. We show here that for a universe populated by
point masses, the entire probability distributions for magnification,
shear, and rotation at each redshift do not depend in an essential way
on either the individual masses or clustering properties of the point
masses. Thus, the point mass probability distributions for
magnification, shear, and rotation are remarkably ``universal'' in
character, depending only upon the choice of underlying
Robertson-Walker model. Our computations of this universal probability
distribution for various choices of $\Omega_0$ and $\Lambda$ will be
given in Subsection~\ref{ss:results}.

It should be emphasized that various lensing effects with null
geodesics which differ by a {\em finite} (as opposed to infinitesimal)
amount {\em will} depend on the masses and clustering properties of
the individual point masses. In particular, if the lensing of a given
source produces multiple images, the expected angular separation of
these images will depend strongly on the masses and/or clustering
properties of the point masses. What we show here, however, is that
for a universe populated by point masses, all of the statistical
properties of the {\em individual images} of nearly point sources are
``universal''---although the criteria for what constitutes a ``nearly
point source'' does depend upon the properties of the point
masses (see Eq.~(\ref{mlim}) above).

\subsection{Random Distributions of Point Masses}
\label{ss:points}

We turn now to a demonstration of claim (1). As in the previous
section, consider the propagation of a beam of photons backwards in
time. We focus attention on the lensing effects caused by matter near
the photon trajectory at redshift $z$. We may view this matter as
lying in a single ``lens plane''. Suppose a point mass of mass
$M$ is randomly placed in this lens plane within a disk of radius $L$
centered on the photon trajectory. By Eqs.~(\ref{A1}) and (\ref{A2}),
the effect of this mass on ${\cal A}^\mu{}_\nu$ is determined by
$J^\mu{}_\nu$. Setting $R \rightarrow \infty$ in Eq.~(\ref{JP}), and
performing the rotation indicated in Eq.~(\ref{Jxx}), we obtain
\begin{eqnarray}
J^{x}{}_{x} &=& - \frac{4 M \cos 2\theta}{b^2} \nonumber \\
J^{y}{}_{y} &=& \ \frac{4 M \cos 2\theta}{b^2} \label{JP'} \\
J^{x}{}_{y} &=& - \frac{4 M \sin 2\theta}{b^2}, \nonumber
\end{eqnarray}
where $(b, \theta)$ (with $b \in [0,L]$)
denote the polar coordinates of the point mass. It follows that the
lensing effects of the point mass on the beam of photons are
uniquely determined via Eq.~(\ref{JP'}) by a knowledge of the probability
distribution, $p(\psi)$, for the variable
\begin{equation}
\psi \equiv \frac{4 M \cos 2 \theta}{b^2},
\label{psi}
\end{equation}
corresponding to a random choice of point in the disk. A
straightforward calculation yields
\begin{equation}
p(\psi) = \frac{4M}{\pi L^2 \psi^2}
\left\{\begin{array}{ll}
1-\sqrt{1-({L^4\psi^2/16M^2})} & \mbox{if $L^4\psi^2/16M^2<1$} \\
1 & \mbox{otherwise}
\end{array} \right. .
\label{ppsi}
\end{equation}
Note that this probability distribution has a divergent second moment
(and the integral defining the first moment fails to converge absolutely).

Now suppose we break up the point mass, $M$, into $N$ point masses,
each of mass $m = M/N$. Suppose we randomly distribute these $N$
point masses within the same disk of radius $L$ in the given lens
plane. The corresponding $J^\mu{}_\nu$ produced by this configuration
of masses is given by a formula similar to Eq.~(\ref{JP'}), except that
$M$ is replaced by $M/N$ and a sum is taken over the independent
contributions of the $N$ particles. By inspection, we see that the
probability distribution for $J^\mu{}_\nu$ in this case is determined
by the probability distribution, $P_N(\Psi)$, for the variable $\Psi$
in exactly the same manner as the probability distribution for
$J^\mu{}_\nu$ in the case of a single mass is determined by $p(\psi)$,
where
\begin{equation}
\Psi \equiv \frac{1}{N} \sum_{i=1}^N \psi_i,
\label{Psi}
\end{equation}
and where each $\psi_i$ is given by Eq.~(\ref{psi}) with $(b, \theta)$
taken to be the polar coordinates of the $i$th particle.  Thus, the
lensing effects of a single, randomly distributed point mass of mass
$M$ will differ from the lensing effects of $N$ randomly distributed
point masses, each of mass $M/N$, precisely to the extent that the
probability distribution functions $p$ and $P_N$ differ.

The determination of the relationship between $p$ and $P_N$ is a
standard problem in probability theory: If $p$ is the probability
distribution for the random variable $\psi$, then $P_N(\Psi)$
corresponds to the probability that the average value of a series of
$N$ independent ``trials'' will be $\Psi$. If $p$ is such that its
second moment is finite, the answer to this problem in the limit of
large $N$ is known as the ``law of large numbers'', which states
that $P_N$ is well
approximated by a Gaussian centered at $<\!\!\psi\!\!>$ whose width is
proportional to $1/\sqrt{N}$. However, the law of large numbers is
{\em not} applicable here because, as already noted above, the
probability distribution of Eq.~(\ref{ppsi}) fails to have a finite second
moment. We note in passing that the failure of $p$ to have a finite
second moment implies that if one wishes to work with moments of $p$,
it will be necessary to impose an ``opaque radius'' cutoff to the
probability distribution at large $\psi$, as has been done in most
analyses of the ``Swiss cheese'' models. No opaque radius cutoff
will be imposed here.

Although the law of large numbers does not hold here, the mathematical
techniques used in the proof of the law of large numbers can be used
to analyze the relationship between $p$ and $P_N$ (see \cite{SEF}).  We write
\begin{equation}
P_N(\Psi) = \int\!\delta\!\left({\Psi - \frac{1}{N}{\textstyle\sum_{i=1}^N}
\psi_i}\right)
p(\psi_1)\cdots p(\psi_N)\,d\psi_1\cdots d\psi_N.
\label{pP}
\end{equation}
Taking the Fourier transform of $P_N$, we obtain
\begin{eqnarray}
\hat{P}_N(K) & \equiv & \int\! e^{-iK\Psi} P_N(\Psi)\,d\Psi \nonumber \\ 
& = & \int\! e^{-iK\psi_1/N}\cdots e^{-iK\psi_N/N}p(\psi_1)\cdots p(\psi_N)
\,d\psi_1 \cdots d\psi_N \nonumber \\
& = & [\hat{p}(K/N)]^N,
\label{Pft}
\end{eqnarray}
where $\hat{p}$ is the Fourier transform of $p$. A direct computation
of $\hat{p}$ from Eq.~(\ref{ppsi}) yields
\begin{equation}
\hat{p}(k) = 1 - 4M |k|/L^2 + O(k^2).
\label{pft}
\end{equation}
In the limit as $N \rightarrow \infty$, we have
\begin{eqnarray}
\hat{P}_\infty (K) & = & \lim_{N \rightarrow \infty}\left({1 -
\frac{4M|K|}{N L^2}}\right)^N \nonumber \\
& = & \exp (- 4M|K|/L^2).
\label{Pftin}
\end{eqnarray}
Taking the inverse Fourier transform of Eq.~(\ref{Pftin}), we obtain
\begin{equation}
P_\infty (\Psi) = \frac{4M}{\pi L^2} \frac{1}{\Psi^2 + \alpha^2},
\label{Pin}
\end{equation}
where $\alpha \equiv 4M/L^2$.

A number of key conclusions follow directly from Eq.~(\ref{Pin}).
First, the mere existence of $P_\infty$ is somewhat surprising, since,
{\it a priori}, there is no obvious reason to expect a well defined,
finite, and nonvanishing limit to the lensing effects of randomly
distributed point masses as $N \rightarrow \infty$. More remarkable
still is the fact that $P_\infty$ nearly coincides with the original
probability distribution $p$; they are both ``$1/x^2$'' distributions
with the same coefficient, but are ``regularized'' in slightly
different ways near $x = 0$. Furthermore, the small difference between
$P_\infty(x)$ and $p(x)$ at $x \approx 0$ is of no importance because
the contribution to lensing is negligible in that regime. Thus, the
probability distributions for magnification, shear, and rotation of
images for the case where a single point mass is randomly placed in a
lens plane will differ negligibly from the limiting case where
infinitely many infinitesimal point masses (with the same total mass)
are randomly placed in that lens plane.

The fact that $P_\infty$ is an excellent approximation to $p$ further
implies that for finite $N$, $P_N$ also can differ only negligibly
from $p$. Indeed, if we approximate $p$ by $P_\infty$ in
Eq.~(\ref{Pft}), we find $P_N \approx P_\infty$, i.e., to the extent
that we have $p \approx P_\infty$ we also have $p \approx P_N$ for all
$N$. Consequently, an arbitrary (finite) subdivision of a point mass
placed randomly in a given lens plane has essentially no effect on the
statistical distributions of lensing images, provided, of course, that
all of the point masses resulting from this subdivision are again
randomly distributed. Since an arbitrary spectrum of masses can be
produced by appropriate subdivisions starting from a single point
mass, this result has the further consequence that the random
placement of {\em any} collection of point masses of total mass $M$ in
a given lens plane has the same effect on the statistical
distributions of the lensing images as a single point mass of mass
$M$. Finally, since the cumulative lensing effects produced by all of
the matter in the universe can be viewed as resulting from a sequence
of encounters of the beam of photons with matter lying in various
different lens planes, we conclude that {\em all} random distributions
of point masses throughout the universe (of {\em any} mass spectrum)
must produce the same statistical distributions of the lensing images,
as stated in claim (1) above.

As a check of both the above arguments and our methods for calculating
lensing effects, we have tested claim (1) in the following ways:
First, we computed the statistical distributions of the lensing images
for a universe with $\Omega = 1$ and $\Lambda = 0$ by the method of
Section~\ref{method}, taking the ``galaxies'' to be point masses of
mass $M = 10^{12} M_\odot$. This should correspond closely to the case
of a universe filled with a random distribution of point masses, each
having mass $10^{12} M_\odot$. Then we repeated the Monte Carlo
calculations choosing $M = 10^{13} M_\odot$, and taking ${\cal R}$ to
be correspondingly larger. The results we obtained for the two cases
were statistically indistinguishable from each other. We also
repeated our calculations with ``galaxies'' taken to be uniform
density balls of radius $d = \cal R$ (so that the matter is distributed
exactly as in the underlying Robertson-Walker model), but taking
``microlensing'' by ``stars'' into account in the manner explained at
the end of Section~\ref{method}. This calculation corresponds to a
universe filled with a random distribution of point masses each of
mass $M = M_\odot$. Again, the results we obtained were statistically
indistinguishable from the two previous cases.

\subsection{Spherical Clustering of Point Masses}
\label{ss:clustering}

The results of the previous subsection apply only to random
distributions of point masses. What happens if the point masses are
clustered? On one hand, since clustering produces corresponding
``voids'', there should be an increased likelihood that the photon
beam will fail to come close to any point mass. Furthermore, if the
photon beam passes near to some point mass, there is an increased
likelihood that the lensing effects of this point mass will be
partially canceled by the presence of other nearby point masses. These
two effects suggest that clustering should decrease the overall
lensing efficiency of point masses. On the other hand, clustering will
result in the production of a large scale, coherent ``cluster
potential'', which can cause important lensing effects on the beam of
photons even if this beam does not pass close to any individual point
mass. This effect suggests that clustering should increase the overall
lensing efficiency of point masses.

In fact, our results for randomly distributed point masses strongly
suggest that the above effects should nearly cancel, at least for
spherical clustering. Imagine starting with a random distribution of
point masses, each of mass $M$. Suppose we cluster $N$ of these point
masses into a spherical structure of radius $d$. This clustering
should have little effect upon lensing unless $d$ is sufficiently
small that the Einstein radii of the individual masses overlap when
projected into the plane perpendicular to the path of the
photon~\cite{dalcanton}. However, at this stage, $d$ will be of the order
of the Einstein radius of the cluster. If $d$ is made still smaller,
the cluster itself can be treated as a ``point mass'', and, by claim
(1), its lensing effects will be equivalent to that of the original
unclustered distribution. Thus, except perhaps for a cluster whose
size is very nearly equal to its own Einstein radius, spherical
clustering of point masses should have a negligible effect on the
statistical distributions of the lensing images, in accord with claim
(2).

We have tested the above arguments for claim (2) by performing a
number of runs for both uniform density and isothermal galaxies of
various radii, using the ``microlensing'' procedure
described at the end of Section~\ref{method}. These calculations
simulate the clustering effects of ``point mass stars'' into
galaxies. We found that the statistical distributions of the lensing
images were indeed statistically indistinguishable from the randomly
distributed point mass case except when the radius of the galaxy was
close to its own Einstein radius. In that case, the clustering
produced a slight---but statistically significant---diminution
of the lensing effects as compared with a random distribution of point
masses. However, even when we chose the parameters of the galaxy so as
to maximize the differences, the effects of clustering were not
significant, as will be illustrated in Fig.~\ref{F:effective1} below.
Thus, to an excellent approximation, for a universe with matter in the
form of point masses, the statistical distributions for the
magnification, shear, and rotation of images is universal in
character, independent of the mass distribution and (spherical)
clustering properties of the point masses.

\subsection{Results}
\label{ss:results}

We now present some of
the results of our Monte Carlo calculations of these distributions for
several cosmological models. 
As already indicated at the beginning of Section~\ref{method}, in our
calculations we used the values $H_0 = 70\ \mbox{km}/\mbox{s Mpc}$
and ${\cal R} = 2\ \mbox{Mpc}$, with the mass, $M$, of each galaxy
then determined from the underlying Robertson-Walker model. Thus, for
a universe with $\Omega = 1$, we set $M=5 \times 10^{12} M_\odot$,
whereas for a universe with $\Omega_0 = 0.1$, we set $M=5 \times
10^{11} M_\odot$. We then performed the calculations described in
detail in Section~\ref{method}, using Eq.~(\ref{JP}) for
$J^\mu{}_\nu$. In each case presented here, we performed 2,000
``runs'' back to a redshift of $3$, and we calculated the
magnification, shear, and rotation of the beam of photons at various
intermediate redshifts as well.

The main focus of our attention was on the distribution of
magnifications, since that distribution is most relevant to the
applications described in Sections~\ref{york} and~\ref{supernovae}
below. We plot our magnification results in the following manner: At
the given redshift of interest, our data set contains 2,000
individual values of area---one for each ``run''. We sort these runs
in order of increasing value of area, with the area of beams which
have passed through a caustic counting as negative (i.e., the large
negative areas come first).\footnote{In some cases, a handful of runs
contained beams which had passed through two caustics (and thus had
(usually very large) positive area). When such double-caustic runs
occurred, we sorted them so that their areas appeared first, i.e.,
before any of the negative areas arising from single caustic runs.}
We then normalize the areas to the area of the underlying
Robertson-Walker model, so that an image corresponding to a beam with
$A=1$ has exactly the same apparent luminosity it would have had in
that model.  In other words, images of objects carried by beams of
photons with $|A| < 1$ have been magnified/amplified with respect to
Robertson-Walker, while those with $|A| > 1$ have been
demagnified/deamplified.

In our figures we plot the area of each run against its position in
the sorted list.  For convenience, we re-scale the $x$-axis so that it
ranges up to 100 rather than 2,000. Thus, for a given area value $A$,
the corresponding $x$-value yields the percentage of beams with area
less than $A$. This ``percentage'' is taken relative to a random
sampling of telescope directions in the present sky, as opposed to a
random sampling of source positions at redshift $z$. Note that if a
source is randomly placed on a sphere of radius $D$ centered on
us (with $D$
chosen so that the light reaching us was emitted at redshift $z$), the
probability that it will be ``hit'' by a given photon beam is
proportional to the magnitude of the area, $|A|$, that the beam has at
redshift $z$.  A large beam will sample a larger section of the sky,
and therefore will represent more sources (all of them demagnified)
than a small beam.  The probability that a randomly placed source will
have an image with area between $A$ and $A + \Delta A$ is proportional
to the corresponding value of $\Delta x \times |A|$, rather than just
$\Delta x$. We will discuss this further in Section~\ref{supernovae}
below (see Eq.~(\ref{pP2})).
\label{area_discussion}

Note also that since lensing simply magnifies or demagnifies images
relative to the underlying Robertson-Walker model---but does not
affect the surface brightness of the images---the apparent luminosity
of an image of a source is proportional to $1/A$. Since, as just noted
above, the probability that a beam ``hits'' a given source is
proportional to $A$, the expected luminosity (i.e., photon flux) in
each beam is exactly the same as in the underlying
Robertson-Walker model. In particular, our analysis automatically
builds in the fact that the expected total luminosity agrees with that
of the underlying Robertson-Walker model.

For beams of photons which have not undergone caustics, the largest
possible area is the ``flat space'' (or ``empty beam''
\cite{dyerroeder2}) value, $A_{\rm Flat}$, corresponding to setting
the curvature to zero in the geodesic deviation equation. This value
is marked on (most of) the figures. It should be noted that after a
photon beam undergoes a caustic, its area typically becomes very large
(and negative)---significantly larger in magnitude than the flat
space value. (An indication of this fact can already be seen in
Fig.~1.) To avoid problems with the scale of our figures, we did not
attempt to plot any area values less than $- A_{\rm Flat}$. This
accounts for the ``gap'' at the beginning of our plots.

We will refer to an image associated with a photon beam which has not
undergone a caustic as a {\it primary image}.\footnote{This
corresponds to the ``type I'' image of~\cite{SEF}.} If $p$ denotes the
event representing our telescope at the present time, then any event
$q$ which lies on the boundary of the past of $p$ must be connected to
$p$ by a null geodesic whose corresponding photon beam has not
undergone a caustic (see, e.g., \cite{wald}). Since the world line of
any source must intersect the boundary of the past of $p$, it follows
that every source must have at least one primary image (see
also~\cite{SEF}). In Subsection~\ref{ss:checks} below we shall
argue that for spherical distributions of matter it is very rare that
a source would have more than one primary image, but for very dense
cylindrical matter distributions, multiple primary images are
common. Every primary image of a source must be at least as bright as
it would be if it were placed in flat spacetime at the same affine
parameter distance~\cite{SEF}. On the other hand, secondary images
(corresponding to photon beams which have undergone one or more
caustics) can be arbitrarily faint. Of course, a secondary image of a
source can also be bright and, in particular, can be brighter than a
primary image of that source.

Each secondary image of a source must have at least one associated
primary image of the same source, and (since the total number of
images must be odd~\cite{SEF}) must also have other associated
secondary images.  If the angular separation of these images is very
small, it may not be possible to resolve the separate images.  One of
the shortcomings of our method is that we do not have a good way of
determining (even statistically) which primary and secondary images
are associated with each other, since this would require us to analyze
photon trajectories which differ by finite (as opposed to
infinitesimal) separations.  Thus, if the different images of the same
source are not resolved---the case of microlensing---we are unable to
predict the probability distribution in total luminosity.

\begin{figure}
\begin{center}
\begin{picture}(0,0)%
\epsfig{file=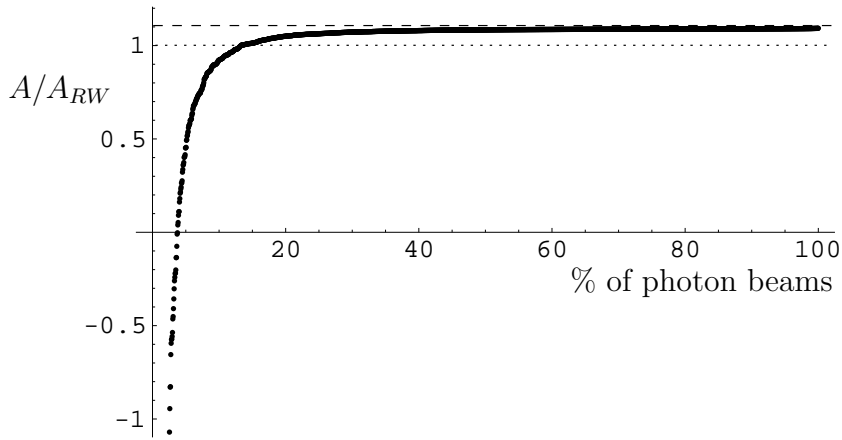}%
\end{picture}%
\setlength{\unitlength}{3947sp}%
\begingroup\makeatletter\ifx\SetFigFont\undefined%
\gdef\SetFigFont#1#2#3#4#5{%
  \reset@font\fontsize{#1}{#2pt}%
  \fontfamily{#3}\fontseries{#4}\fontshape{#5}%
  \selectfont}%
\fi\endgroup%
\begin{picture}(5238,2742)(4360,-3691)
\put(7887,-2753){\makebox(0,0)[lb]{\smash{\SetFigFont{12}{14.4}{\familydefault}{\mddefault}{\updefault}
\% of photon beams
}}}
\put(4360,-1538){\makebox(0,0)[lb]{\smash{\SetFigFont{12}{14.4}{\familydefault}{\mddefault}{\updefault}
$A/A_{RW}$
}}}
\end{picture}
\end{center}
\caption{Area vs. \% of photon beams at $z=0.5$, for an $\Omega=1$,
$\Lambda=0$ universe, with matter distributed in the form of point masses.
The dashed line represents the flat spacetime (empty beam) area,
and the dotted line represents the Robertson-Walker area.}
\label{F:z_05}
\end{figure}

\begin{figure}
\begin{center}
\begin{picture}(0,0)%
\epsfig{file=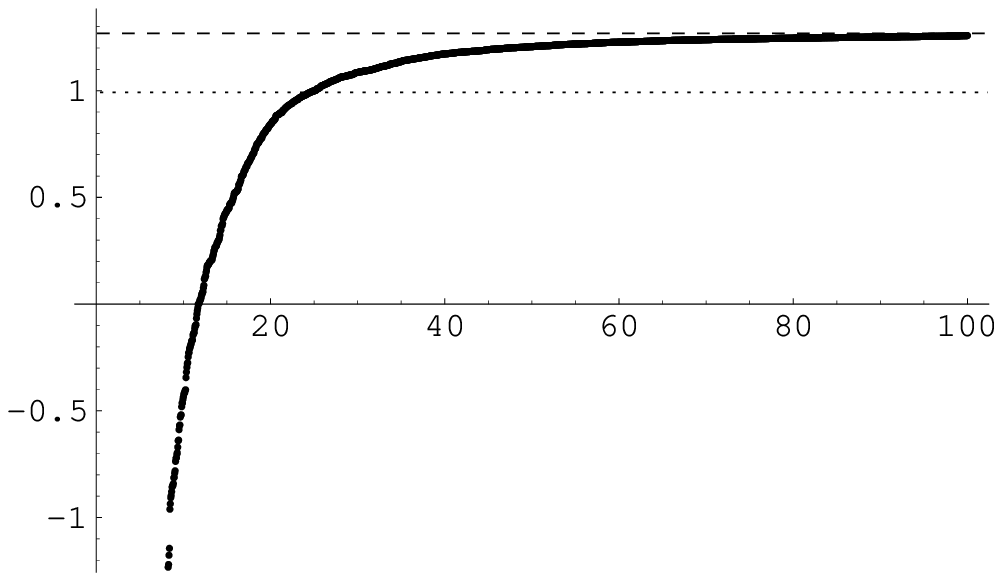}%
\end{picture}%
\setlength{\unitlength}{3947sp}%
\begingroup\makeatletter\ifx\SetFigFont\undefined%
\gdef\SetFigFont#1#2#3#4#5{%
  \reset@font\fontsize{#1}{#2pt}%
  \fontfamily{#3}\fontseries{#4}\fontshape{#5}%
  \selectfont}%
\fi\endgroup%
\begin{picture}(5239,2742)(4359,-3691)
\put(7878,-2760){\makebox(0,0)[lb]{\smash{\SetFigFont{12}{14.4}{\familydefault}{\mddefault}{\updefault}
\% of photon beams
}}}
\put(4359,-1674){\makebox(0,0)[lb]{\smash{\SetFigFont{12}{14.4}{\familydefault}{\mddefault}{\updefault}
$A/A_{RW}$
}}}
\end{picture}
\end{center}
\caption{Area vs. \% of photon beams at $z=1.0$, for an $\Omega=1$,
$\Lambda=0$ universe. Matter is distributed in the form of point masses.}
\label{F:z_1}
\end{figure}

\begin{figure}
\begin{center}
\begin{picture}(0,0)%
\epsfig{file=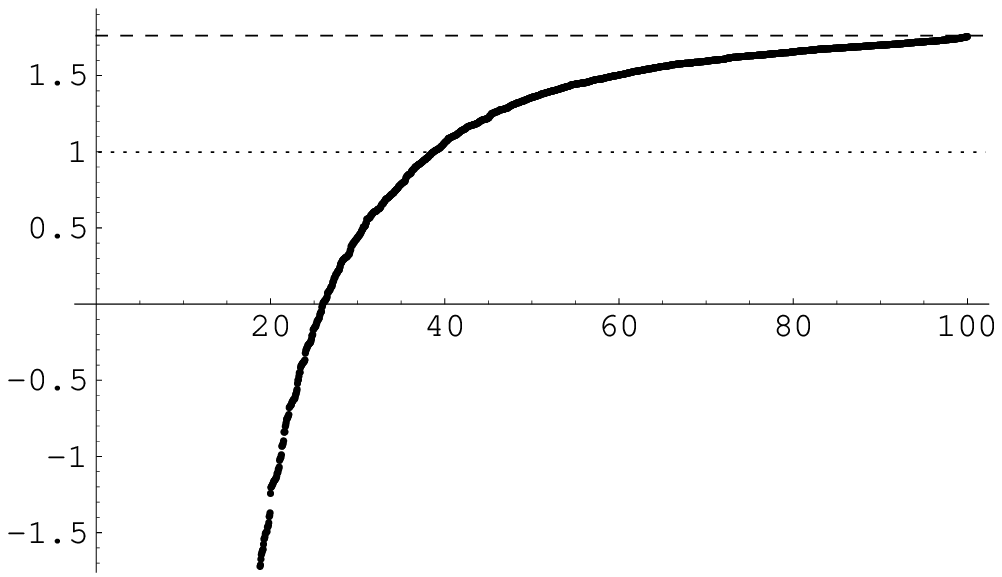}%
\end{picture}%
\setlength{\unitlength}{3947sp}%
\begingroup\makeatletter\ifx\SetFigFont\undefined%
\gdef\SetFigFont#1#2#3#4#5{%
  \reset@font\fontsize{#1}{#2pt}%
  \fontfamily{#3}\fontseries{#4}\fontshape{#5}%
  \selectfont}%
\fi\endgroup%
\begin{picture}(5239,2742)(4359,-3691)
\put(4359,-1714){\makebox(0,0)[lb]{\smash{\SetFigFont{12}{14.4}{\familydefault}{\mddefault}{\updefault}
$A/A_{RW}$
}}}
\put(7870,-2753){\makebox(0,0)[lb]{\smash{\SetFigFont{12}{14.4}{\familydefault}{\mddefault}{\updefault}
\% of photon beams
}}}
\end{picture}
\end{center}
\caption{Area vs. \% of photon beams at $z=2.0$, for an $\Omega=1$,
$\Lambda=0$ universe. Matter is distributed in the form of point masses.}
\label{F:z_2}
\end{figure}

\begin{figure}
\begin{center}
\begin{picture}(0,0)%
\epsfig{file=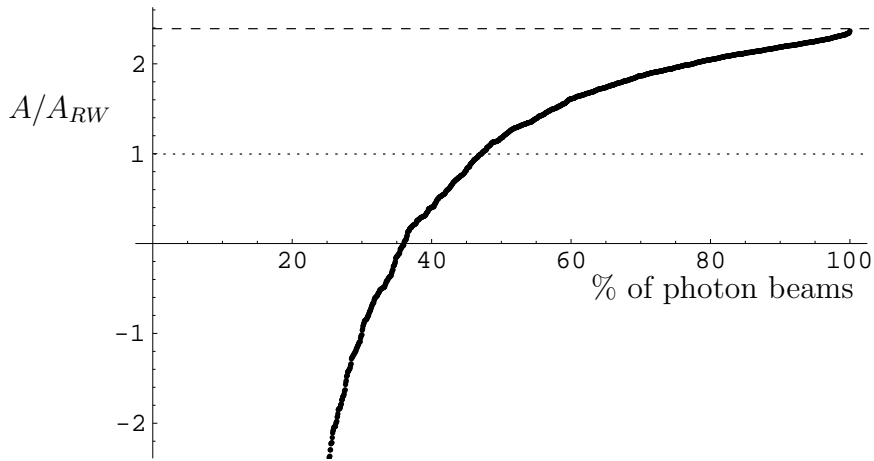}%
\end{picture}%
\setlength{\unitlength}{3947sp}%
\begingroup\makeatletter\ifx\SetFigFont\undefined%
\gdef\SetFigFont#1#2#3#4#5{%
  \reset@font\fontsize{#1}{#2pt}%
  \fontfamily{#3}\fontseries{#4}\fontshape{#5}%
  \selectfont}%
\fi\endgroup%
\begin{picture}(5442,2876)(4156,-3825)
\put(4156,-1660){\makebox(0,0)[lb]{\smash{\SetFigFont{12}{14.4}{\familydefault}{\mddefault}{\updefault}
$A/A_{RW}$
}}}
\put(7816,-2793){\makebox(0,0)[lb]{\smash{\SetFigFont{12}{14.4}{\familydefault}{\mddefault}{\updefault}
\% of photon beams
}}}
\end{picture}
\end{center}
\caption{Area vs. \% of photon beams at $z=3.0$, for an $\Omega=1$,
$\Lambda=0$ universe. Matter is distributed in the form of point masses.}
\label{F:z_3}
\end{figure}

Figs.~\ref{F:z_05}--\ref{F:z_3} show our results for a universe filled
with point masses corresponding to an underlying spatially flat
Robertson-Walker cosmology with $\Omega = 1$ and $\Lambda = 0$. As
with all plots shown here, we have taken $H_0 = 70\
\mbox{km}\,\mbox{s}^{-1}\,\mbox{Mpc}^{-1}$. We also took ${\cal R} =
2\ \mbox{Mpc}$, although as argued above, the results should be
independent of the choice of ${\cal R}$. As can be seen from the
graphs, the percentage of photon beams which have undergone caustics
ranges from about $5\%$ at redshift $z = 1/2$ to over $35\%$ at
redshift $z = 3$. Note also that by redshift $z=3$ about $20\%$ of the
primary images are less than half as bright ($A > 2$) as they would have
been in in the underlying Robertson-Walker cosmology.  Taking account
of the factor of $|A|$ mentioned above, and assuming that each source
has only one primary image (see Subsec.~\ref{ss:checks} below),
we find that the probability that the primary image of a randomly
placed source at $z=3$ will be demagnified relative to Robertson-Walker by at
least a factor of $2$ is $1/2$.  Since these photon trajectories do
not pass near any of the point masses, it seems unlikely that such
sources will have any (bright) secondary images. Thus, even if
multiple images cannot be resolved, it appears that in this cosmology,
at redshift $z=3$, $50\%$ of all sources should be dimmer by at least
a factor of $2$ relative to the underlying Robertson-Walker model.

\begin{figure}
\begin{center}
\begin{picture}(0,0)%
\epsfig{file=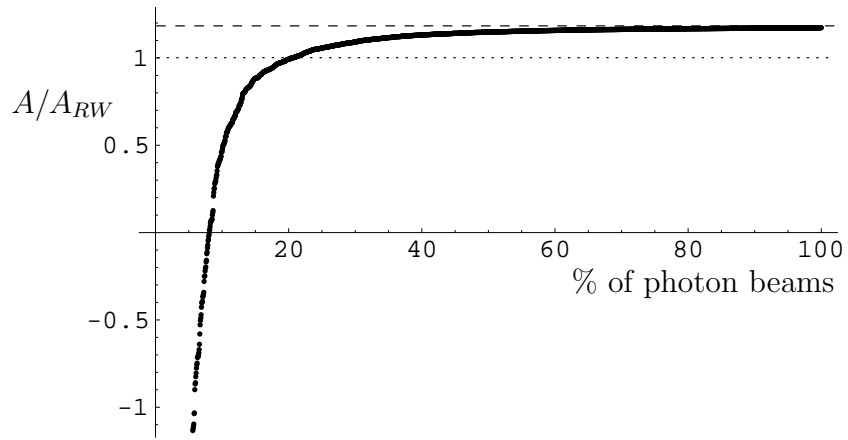}%
\end{picture}%
\setlength{\unitlength}{3947sp}%
\begingroup\makeatletter\ifx\SetFigFont\undefined%
\gdef\SetFigFont#1#2#3#4#5{%
  \reset@font\fontsize{#1}{#2pt}%
  \fontfamily{#3}\fontseries{#4}\fontshape{#5}%
  \selectfont}%
\fi\endgroup%
\begin{picture}(5239,2742)(4359,-3691)
\put(4359,-1627){\makebox(0,0)[lb]{\smash{\SetFigFont{12}{14.4}{\familydefault}{\mddefault}{\updefault}
$A/A_{RW}$
}}}
\put(7878,-2753){\makebox(0,0)[lb]{\smash{\SetFigFont{12}{14.4}{\familydefault}{\mddefault}{\updefault}
\% of photon beams
}}}
\end{picture}
\end{center}
\caption{Area vs. \% of photon beams at $z=3.0$, for an $\Omega_0=0.1$,
$\Lambda=0$ universe. Matter is distributed in the form of point masses.}
\label{F:om01}
\end{figure}

The results at $z = 3$ for a universe filled with
point masses corresponding to an open Robertson-Walker model 
with $\Omega_0=0.1$ and $\Lambda=0$ is plotted in Fig.~\ref{F:om01}.
It can be seen that the lensing effects here are dramatically weaker
than in the $\Omega=1$ model. In particular, 
in this cosmology less than $10\%$ of the photon beams have
undergone a caustic by $z=3$, and the maximum de-magnification relative to
Robertson-Walker is only $0.85$ (but over half the primary images suffer
nearly this de-magnification).

\begin{figure}
\begin{center}
\begin{picture}(0,0)%
\epsfig{file=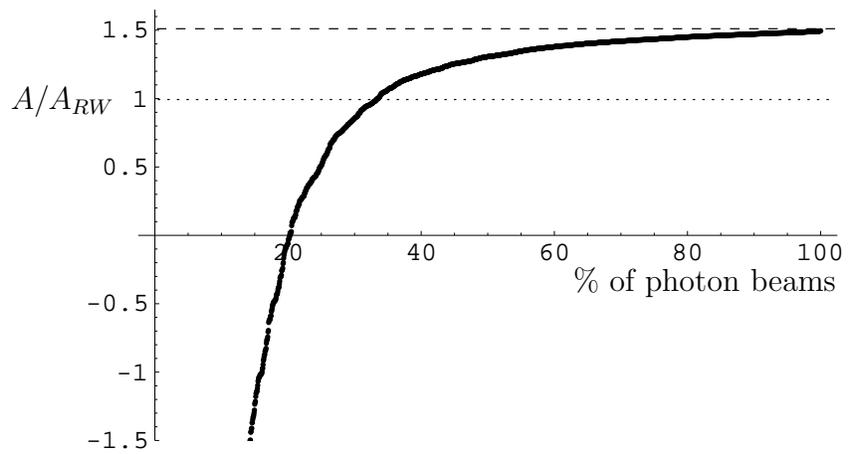}%
\end{picture}%
\setlength{\unitlength}{3947sp}%
\begingroup\makeatletter\ifx\SetFigFont\undefined%
\gdef\SetFigFont#1#2#3#4#5{%
  \reset@font\fontsize{#1}{#2pt}%
  \fontfamily{#3}\fontseries{#4}\fontshape{#5}%
  \selectfont}%
\fi\endgroup%
\begin{picture}(5246,2809)(4352,-3158)
\put(4352,-985){\makebox(0,0)[lb]{\smash{\SetFigFont{12}{14.4}{\familydefault}{\mddefault}{\updefault}
$A/A_{RW}$
}}}
\put(7884,-2131){\makebox(0,0)[lb]{\smash{\SetFigFont{12}{14.4}{\familydefault}{\mddefault}{\updefault}
\% of photon beams
}}}
\end{picture}
\end{center}
\caption{Area vs. \% of photon beams at $z=3.0$, for an $\Omega_0=0.1$,
$\Omega_\Lambda=0.9$ universe. Matter is distributed in the form of point
masses.}
\label{F:om01la09}
\end{figure}

Finally, the results at $z = 3$ for a universe filled with point
masses corresponding to a spatially flat Robertson-Walker model with
$\Omega_0=0.1$ and $\Omega_\Lambda \equiv \Lambda/3{H_0}^2 =0.9$ are
plotted in Fig.~\ref{F:om01la09}. This distribution is intermediate
between the cases of $\Omega=1$, $\Lambda=0$ and $\Omega_0=0.1$,
$\Lambda=0$.

\begin{figure}
\begin{center}
\begin{picture}(0,0)%
\epsfig{file=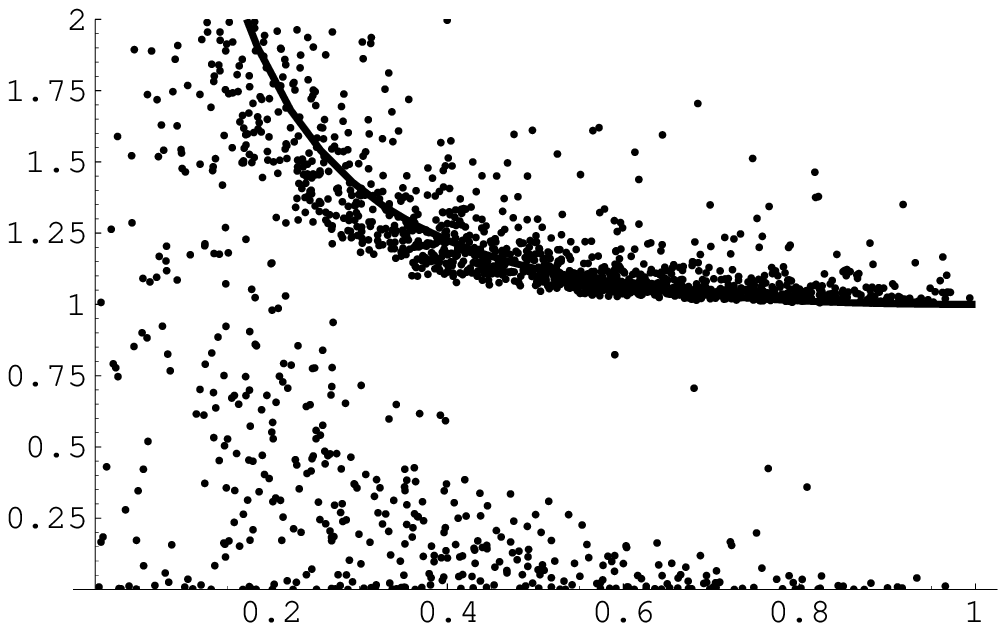}%
\end{picture}%
\setlength{\unitlength}{3947sp}%
\begingroup\makeatletter\ifx\SetFigFont\undefined%
\gdef\SetFigFont#1#2#3#4#5{%
  \reset@font\fontsize{#1}{#2pt}%
  \fontfamily{#3}\fontseries{#4}\fontshape{#5}%
  \selectfont}%
\fi\endgroup%
\begin{picture}(5029,3219)(3969,-4750)
\put(6751,-4711){\makebox(0,0)[lb]{\smash{\SetFigFont{12}{14.4}{\familydefault}{\mddefault}{\updefault}
$\epsilon$
}}}
\put(3969,-3015){\makebox(0,0)[lb]{\smash{\SetFigFont{12}{14.4}{\familydefault}{\mddefault}{\updefault}
$\mu$
}}}
\end{picture}
\end{center}
\caption{Magnification vs. axial ratio at $z = 2$, for an $\Omega = 1$,
$\Lambda=0$ universe filled with point masses. The solid
line gives the fit $\mu=(1+\epsilon)^2/(4\epsilon)$, which would hold
if the lensing was done by a single point mass, as described
in~\protect\cite{SEF}. This figure compares well with Fig.~11.12 of that
reference.}
\label{F:shear1}
\end{figure}

A sample of our results for shear is given in Fig.~\ref{F:shear1}. Here
we have plotted the magnification, $\mu$, relative to the empty beam value,
versus the axial ratio, $\epsilon$, of the beam at redshift
$z = 2$ for a universe filled with point masses for the case $\Omega=1$
and $\Lambda=0$. This figure corresponds to Fig.~11.12 of \cite{SEF}, except 
that we also have included the points with $\mu < 1$, arising from
beams which have undergone caustics. The agreement between the figures
appears to be excellent.

\begin{figure}
\begin{center}
\begin{picture}(0,0)%
\epsfig{file=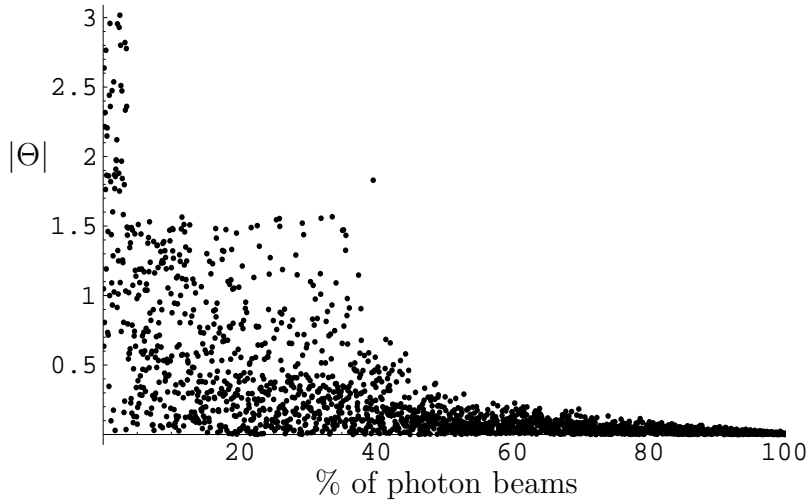}%
\end{picture}%
\setlength{\unitlength}{3947sp}%
\begingroup\makeatletter\ifx\SetFigFont\undefined%
\gdef\SetFigFont#1#2#3#4#5{%
  \reset@font\fontsize{#1}{#2pt}%
  \fontfamily{#3}\fontseries{#4}\fontshape{#5}%
  \selectfont}%
\fi\endgroup%
\begin{picture}(5044,3108)(3954,-4282)
\put(3954,-2183){\makebox(0,0)[lb]{\smash{\SetFigFont{12}{14.4}{\familydefault}{\mddefault}{\updefault}
$|\Theta|$
}}}
\put(5889,-4246){\makebox(0,0)[lb]{\smash{\SetFigFont{12}{14.4}{\familydefault}{\mddefault}{\updefault}
\% of photon beams
}}}
\end{picture}
\end{center}
\caption{Magnitude of rotation angle vs. \% of photon beams at
redshift $z = 3$, for an $\Omega=1$, $\Lambda=0$
universe filled with point masses. The demarcation between beams which
have undergone caustics and those which have not occurs at 36\% (see
Fig.~\ref{F:z_3}). The restriction of $|\Theta|$ to the range $0$ to
$\pi/2$ (rather than $0$ to $\pi$) for beams which have undergone a
single caustic is due to our convention in the definition of $\Theta$
in that case, as explained below Eq.~(\ref{shear}). The first 3\% of
the photon beams have undergone two caustics.}
\label{F:rotation1}
\end{figure}

A sample of our results for rotation is given in
Fig.~\ref{F:rotation1}. Here we plot the magnitude of rotation angle,
$|\Theta|$ (in radians), versus photon beam number (ordered by area,
as described above) at redshift $z = 3$ for a universe filled with
point masses for the case $\Omega=1$ and $\Lambda=0$. From the
figure it can be seen that the photon beams which have not undergone
caustics generally have a very small rotation, but those which have
undergone caustics have undergone such a large rotation that their
orientation is practically random (see \cite{ehlers} for a general
discussion of the behavior of beams near caustics). As noted above, no
rotation would occur for lensing produced by a single point mass.

\begin{figure}
\begin{center}
\begin{picture}(0,0)%
\epsfig{file=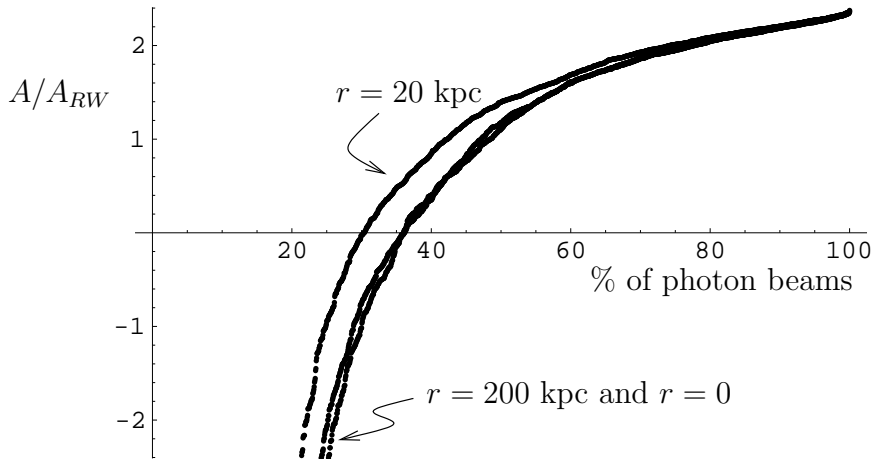}%
\end{picture}%
\setlength{\unitlength}{3947sp}%
\begingroup\makeatletter\ifx\SetFigFont\undefined%
\gdef\SetFigFont#1#2#3#4#5{%
  \reset@font\fontsize{#1}{#2pt}%
  \fontfamily{#3}\fontseries{#4}\fontshape{#5}%
  \selectfont}%
\fi\endgroup%
\begin{picture}(5445,2876)(3553,-3825)
\put(5611,-1561){\makebox(0,0)[lb]{\smash{\SetFigFont{12}{14.4}{\familydefault}{\mddefault}{\updefault}
$r=20$ kpc
}}}
\put(7209,-2746){\makebox(0,0)[lb]{\smash{\SetFigFont{12}{14.4}{\familydefault}{\mddefault}{\updefault}
\% of photon beams
}}}
\put(3553,-1562){\makebox(0,0)[lb]{\smash{\SetFigFont{12}{14.4}{\familydefault}{\mddefault}{\updefault}
$A/A_{RW}$
}}}
\put(6181,-3451){\makebox(0,0)[lb]{\smash{\SetFigFont{12}{14.4}{\familydefault}{\mddefault}{\updefault}
$r=200\ \mbox{kpc}$ and $r=0$
}}}
\end{picture}
\end{center}
\caption{Area vs. \% of photon beams at $z = 3$,
for an $\Omega=1$, $\Lambda=0$ universe.
Graphs are shown for point mass galaxies ($r=0$), and
for uniform density spheres of $r=20\ \mbox{kpc}$ and
$r=200\ \mbox{kpc}$, each composed of (point mass) stars.}
\label{F:effective1}
\end{figure}

Finally Fig.~\ref{F:effective1} shows how remarkably small the effects
of clustering are. The right-most curve shows the magnification versus
photon beam number for point mass galaxies in a universe with
$\Omega=1$ and $\Lambda=0$, at a redshift of 3; it is the same curve
as shown in Fig.~\ref{F:z_3} above. Also shown is the curve for (point
mass) stars clustered into uniform density galaxies of radius $200\
\mbox{kpc}$. This curve is statistically indistinguishable from
the curve for point mass galaxies.
The left-most curve is for
(point mass) stars clustered into uniform density galaxies of radius
$20\ \mbox{kpc}$. This clustering distribution was chosen (in a
parameter search, varying the galactic radius) so as to {\em maximize}
the deviation from the random distribution.  As expected, the maximum
deviation occurs for galaxies (composed of point mass stars) whose
radii are close to their Einstein radii.  It can be seen from the
figure that there is a slight (but statistically significant)
diminution of the lensing effectiveness due to the clustering.

\subsection{A Conjecture}
\label{ss:conjecture}

We conclude this section with a conjecture, based upon the fact that
spherical clustering of point masses appears to
slightly reduce their lensing effectiveness, together with
our expectation (borne out in all of
our simulations) that point masses are more effective in lensing than
any bodies of finite extent:

{\it Conjecture:} For any underlying Robertson-Walker cosmological model
at any redshift $z$, randomly distributed point masses provide the
most ``effective'' distribution of matter
for lensing in the following sense: Let
$A_{\rm rpm}(x)$ denote the area as a function of the percentage of 
photon beams for a universe filled with a
random distribution of point masses (see 
Figs.~\ref{F:z_05}--\ref{F:om01la09}). Let $x_1$ denote the $x$-value
such that $A_{\rm rpm}$ equals the Robertson-Walker area, i.e.,
$A_{\rm rpm}(x_1) = 1$. Then for any other matter distribution, we have
$A(x) > A_{\rm rpm}(x)$ for all $x \leq x_1$. In particular, the greatest
number of caustics is achieved for the case of randomly distributed 
point masses.

\section{Other Cases; Consistency Checks}
\label{cases_and_checks}

\subsection{Uniform Density Balls, Isothermal Balls, and Uniform
Density Cylinders}
\label{ss:cases}

In the previous section the lensing effects occurring in a universe
filled with point masses were analyzed in detail. In this subsection
the corresponding lensing effects will be briefly discussed for other
galactic mass distributions, specifically for isothermal balls,
uniform density balls, and uniform density cylinders. The isothermal
balls should be good descriptions of the actual mass distribution in
galaxies, so (with appropriate choices of parameters) the results for
this case should provide a realistic description of the statistical
lensing effects occurring in our universe---at least provided that
the clustering of galaxies does not play an important role and that
the effects of sub-galactic structure can be ignored. (As previously
discussed near Eq.~(\ref{mlim2}) above, sub-galactic structure need
not be taken into account if the sources are of sufficiently large
angular size. If the sources are of sufficiently small angular size
that ``microlensing'' by stars is of relevance, and if most of the
matter is in the form of stars or other condensed objects, then the
point mass results of the previous section should apply.) Although the
uniform density balls and cylinders presumably do not correspond to
realistic mass distributions,\footnote{Recent findings of Meiksin
{\em et al.}  \cite{meiksin} and others indicate that filamentary
structures may play an important role in the evolution of structure.
Cylindrical mass distributions can be used to mimic filaments, and
identify qualitative differences in the lensing distributions to be
expected from such structures.} they provide useful ``toy models'' for
investigating various effects.

\begin{figure}
\begin{center}
\begin{picture}(0,0)%
\epsfig{file=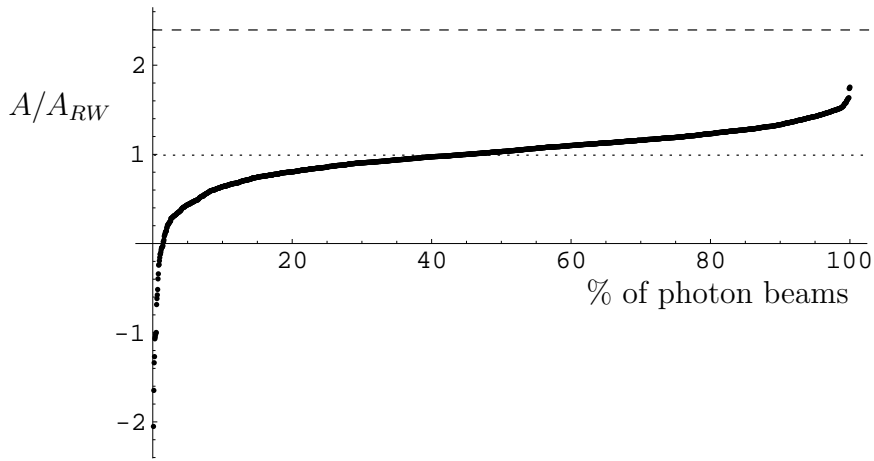}%
\end{picture}%
\setlength{\unitlength}{3947sp}%
\begingroup\makeatletter\ifx\SetFigFont\undefined%
\gdef\SetFigFont#1#2#3#4#5{%
  \reset@font\fontsize{#1}{#2pt}%
  \fontfamily{#3}\fontseries{#4}\fontshape{#5}%
  \selectfont}%
\fi\endgroup%
\begin{picture}(5442,2876)(4156,-3825)
\put(4156,-1645){\makebox(0,0)[lb]{\smash{\SetFigFont{12}{14.4}{\familydefault}{\mddefault}{\updefault}
$A/A_{RW}$
}}}
\put(7785,-2823){\makebox(0,0)[lb]{\smash{\SetFigFont{12}{14.4}{\familydefault}{\mddefault}{\updefault}
\% of photon beams
}}}
\end{picture}
\end{center}
\caption{Area vs. \% of photon beams at $z=3.0$, for an $\Omega=1$,
$\Lambda=0$ universe. Matter is distributed in isothermal ($M\propto r$)
balls of radius $200\ \mbox{kpc}$.}
\label{F:iso200}
\end{figure}

\begin{figure}
\begin{center}
\begin{picture}(0,0)%
\epsfig{file=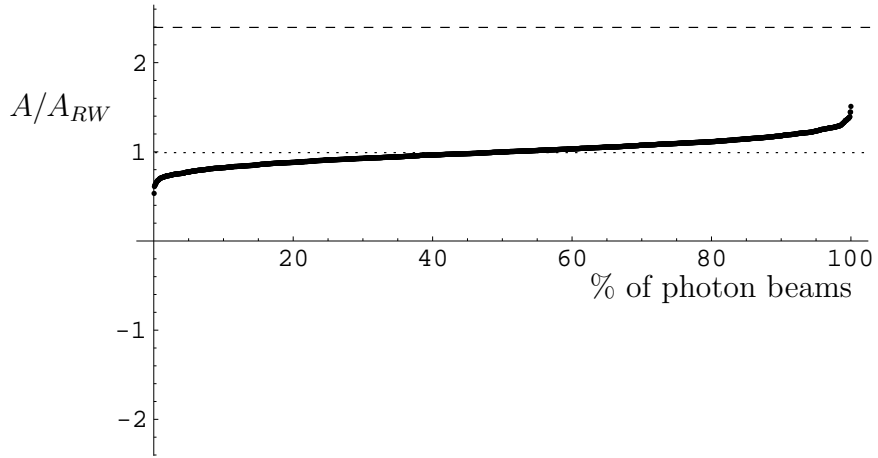}%
\end{picture}%
\setlength{\unitlength}{3947sp}%
\begingroup\makeatletter\ifx\SetFigFont\undefined%
\gdef\SetFigFont#1#2#3#4#5{%
  \reset@font\fontsize{#1}{#2pt}%
  \fontfamily{#3}\fontseries{#4}\fontshape{#5}%
  \selectfont}%
\fi\endgroup%
\begin{picture}(5442,2876)(4156,-3825)
\put(4156,-1667){\makebox(0,0)[lb]{\smash{\SetFigFont{12}{14.4}{\familydefault}{\mddefault}{\updefault}
$A/A_{RW}$
}}}
\put(7801,-2800){\makebox(0,0)[lb]{\smash{\SetFigFont{12}{14.4}{\familydefault}{\mddefault}{\updefault}
\% of photon beams
}}}
\end{picture}
\end{center}
\caption{Area vs. \% of photon beams at $z=3.0$, for an $\Omega=1$,
$\Lambda=0$ universe. Matter is distributed in uniform density
balls of radius $200\ \mbox{kpc}$.}
\label{F:uni200}
\end{figure}

\begin{figure}
\begin{center}
\begin{picture}(0,0)%
\epsfig{file=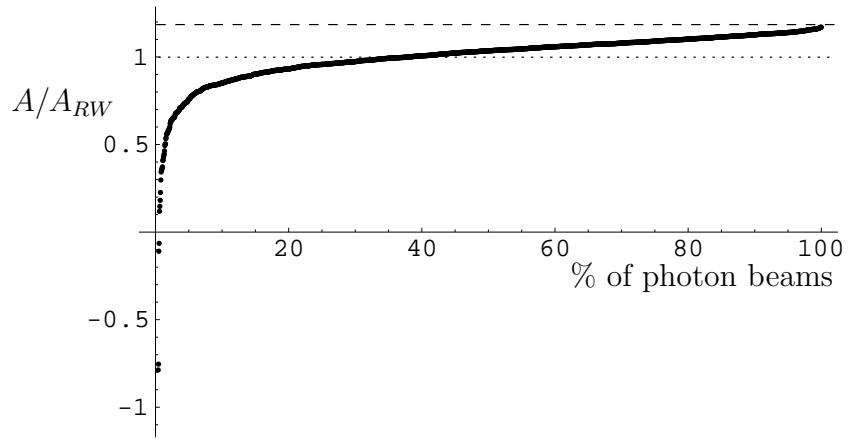}%
\end{picture}%
\setlength{\unitlength}{3947sp}%
\begingroup\makeatletter\ifx\SetFigFont\undefined%
\gdef\SetFigFont#1#2#3#4#5{%
  \reset@font\fontsize{#1}{#2pt}%
  \fontfamily{#3}\fontseries{#4}\fontshape{#5}%
  \selectfont}%
\fi\endgroup%
\begin{picture}(5238,2742)(4360,-3691)
\put(4360,-1618){\makebox(0,0)[lb]{\smash{\SetFigFont{12}{14.4}{\familydefault}{\mddefault}{\updefault}
$A/A_{RW}$
}}}
\put(7870,-2721){\makebox(0,0)[lb]{\smash{\SetFigFont{12}{14.4}{\familydefault}{\mddefault}{\updefault}
\% of photon beams
}}}
\end{picture}
\end{center}
\caption{Area vs. \% of photon beams, at $z=3.0$, for an $\Omega_0=0.1$,
$\Lambda=0$ universe. Matter is distributed in isothermal
balls of radius $50\ \mbox{kpc}$.}
\label{F:iso50}
\end{figure}

\begin{figure}
\begin{center}
\begin{picture}(0,0)%
\epsfig{file=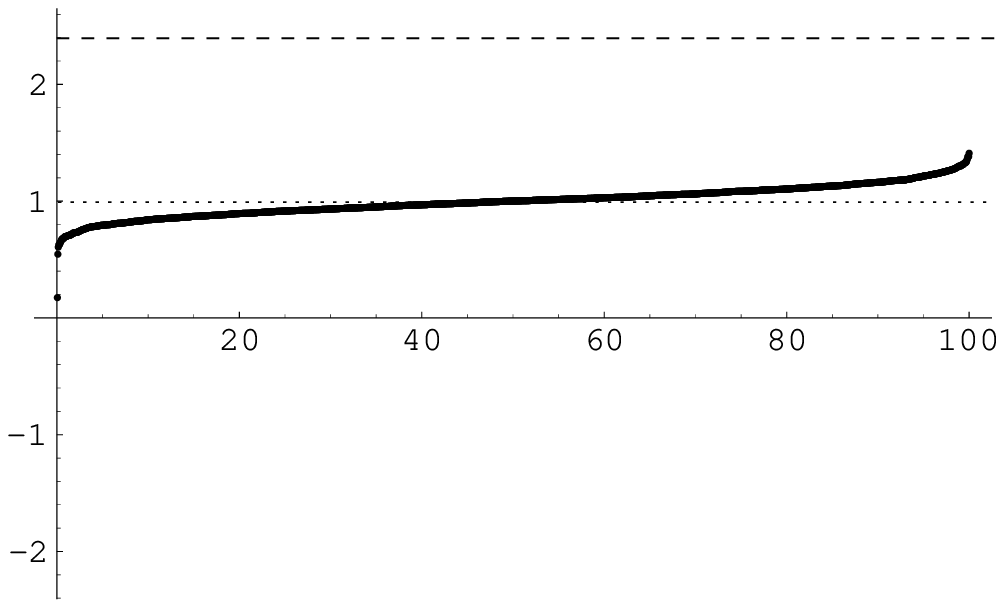}%
\end{picture}%
\setlength{\unitlength}{3947sp}%
\begingroup\makeatletter\ifx\SetFigFont\undefined%
\gdef\SetFigFont#1#2#3#4#5{%
  \reset@font\fontsize{#1}{#2pt}%
  \fontfamily{#3}\fontseries{#4}\fontshape{#5}%
  \selectfont}%
\fi\endgroup%
\begin{picture}(5442,2876)(4156,-3825)
\put(7800,-2808){\makebox(0,0)[lb]{\smash{\SetFigFont{12}{14.4}{\familydefault}{\mddefault}{\updefault}
\% of photon beams
}}}
\put(4156,-1652){\makebox(0,0)[lb]{\smash{\SetFigFont{12}{14.4}{\familydefault}{\mddefault}{\updefault}
$A/A_{RW}$
}}}
\end{picture}
\end{center}
\caption{Area vs. \% of photon beams at $z=3.0$, for an $\Omega=1$,
$\Lambda=0$ universe. Matter is distributed in uniform density cylinders of
radius $52\ \mbox{kpc}$. This figure is almost identical to
Fig.~\ref{F:uni200}, which is for uniform density spheres of the same
density as these cylinders.}
\label{F:cyl52}
\end{figure}

\begin{figure}
\begin{center}
\begin{picture}(0,0)%
\epsfig{file=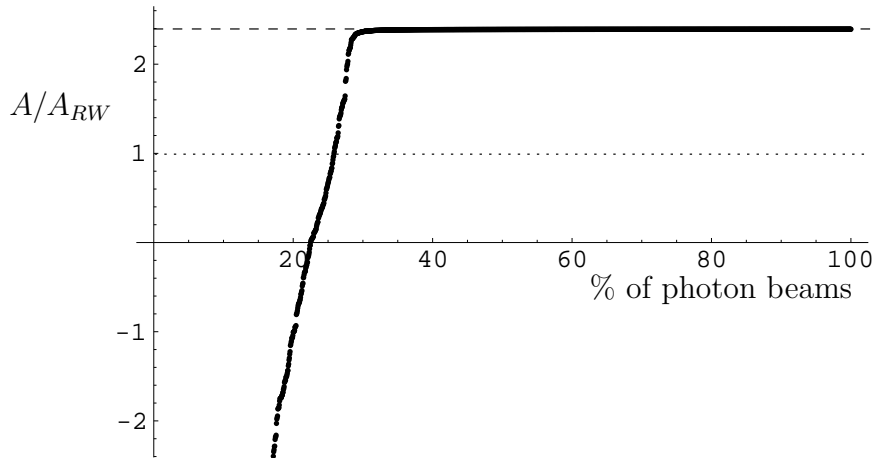}%
\end{picture}%
\setlength{\unitlength}{3947sp}%
\begingroup\makeatletter\ifx\SetFigFont\undefined%
\gdef\SetFigFont#1#2#3#4#5{%
  \reset@font\fontsize{#1}{#2pt}%
  \fontfamily{#3}\fontseries{#4}\fontshape{#5}%
  \selectfont}%
\fi\endgroup%
\begin{picture}(5442,2876)(4156,-3825)
\put(4156,-1653){\makebox(0,0)[lb]{\smash{\SetFigFont{12}{14.4}{\familydefault}{\mddefault}{\updefault}
$A/A_{RW}$
}}}
\put(7801,-2800){\makebox(0,0)[lb]{\smash{\SetFigFont{12}{14.4}{\familydefault}{\mddefault}{\updefault}
\% of photon beams
}}}
\end{picture}
\end{center}
\caption{Area vs. \% of photon beams at $z=3.0$, for an $\Omega=1$,
$\Lambda=0$ universe. Matter is distributed in uniform density
cylinders of radius $200\ \mbox{pc}$.}
\label{F:cyl02}
\end{figure}

As in the point mass case, for the calculations of this subsection we
used the values $H_0 = 70\ \mbox{km}\,\mbox{s}^{-1}\,\mbox{Mpc}^{-1}$
and ${\cal R} = 2\ \mbox{Mpc}$, and we determined the mass, $M$, of
each galaxy from the underlying Robertson-Walker model. Indeed, the
only important difference in our calculational procedures from those of the
point mass case was our use of Eqs.~(\ref{JU}),~(\ref{JI}),
or~(\ref{JC}) in place of Eq.~(\ref{JP}). In addition, in the cylindrical
case, a further randomization over the orientation of the cylinder was
performed at each step. A sampling of some of our results is presented
in Figs.~\ref{F:iso200}-\ref{F:cyl02}.

In Fig.~\ref{F:iso200} we plot the area vs. $\%$ of photon beams at
$z=3$ for a universe with $\Omega=1$ and $\Lambda=0$, populated by
galaxies with a ``truncated isothermal profile'' ($M(r) \propto r$)
and a cutoff radius of $200\ \mbox{kpc}$.  Comparison with
Fig.~\ref{F:z_3} immediately shows that the lensing effects are
greatly reduced as compared with the point mass case. In particular,
only about $2 \%$ of the photon beams have undergone caustics by a
redshift of $3$, as compared with over $35 \%$ in the point mass
case. The lensing effectiveness is further reduced if the galaxies are
modeled as uniform density--rather than isothermal--balls of radius
$200\ \mbox{kpc}$ (see Fig.~\ref{F:uni200}).

Fig.~\ref{F:iso50} shows the area vs. \% of photon beams at $z=3$
for the case $\Omega_0=0.1$ and $\Lambda=0$, with matter distributed in
isothermal balls, now of radius of $50\ \mbox{kpc}$. Again, a significant
reduction in the lensing effectiveness as compared with the point mass
case can be seen (see Fig.~\ref{F:om01}).

Finally, Figs.~\ref{F:cyl52} and~\ref{F:cyl02} plot the area vs. $\%$
of photon beams at $z=3$ for the case $\Omega=1$ and $\Lambda=0$,
with matter distributed in the form of uniform density cylinders of
length $2{\cal R}$ and
cylindrical radii $52\ \mbox{kpc}$ and $200\ \mbox{pc}$,
respectively. As will be discussed further below, the results shown in
Fig.~\ref{F:cyl02} are qualitatively different from all of the other
cases shown here.

\subsection{Consistency Checks}
\label{ss:checks}

The results presented thus far allow us to
perform a number of consistency checks, which provide a good test of
some of the underlying ideas of Section~\ref{intro} as well as a good
test of the validity and accuracy of our formulas and calculational
procedures.

First, as noted in Subsection~\ref{ss:relevant_scales}, the effects of
Ricci curvature on lensing should depend only upon the density
contrasts present in the universe. Thus, in situations where the Weyl
curvature can be neglected, galaxies of the same density should
produce identical statistical lensing effects, independent of their
``shapes''.  This is borne out by a comparison of Figs.~\ref{F:uni200}
and~\ref{F:cyl52}, corresponding, respectively, to spherical and
cylindrical galaxies of the same density. No statistically significant
difference is discernible for the area distributions in these two
cases.

A further important consistency check arises from the fact that the
causal structure of the spacetime of Eq.~(\ref{metric})---whose
properties we are attempting to model with our Monte Carlo
calculations---is nearly the same as that of the underlying
Robertson-Walker model. Consequently, at any redshift $z$, the area of
the boundary of the past of an event $p$ in the spacetime of
Eq.~(\ref{metric}) should be very nearly equal to the area of the past
light cone of $p$ at redshift $z$ in the underlying Robertson-Walker
model. Now, as previously mentioned in Subsection~\ref{ss:results},
for a null geodesic to lie on the boundary of the past of $p$, it is
necessary that its corresponding photon beam not have undergone a
caustic. Thus, in any of our Monte
Carlo results, if we add up the areas of all of the photon beams
which have not undergone caustics (corresponding to ``primary images''
in the terminology introduced in Subsec.~\ref{ss:results}), the
result should be at least as large as the area of the past light cone
in the underlying Robertson-Walker model. Since we have normalized our
beam areas so that $A=1$ corresponds to the Robertson-Walker value,
this means that if we do $N$ Monte-Carlo runs, then, within
statistics, we always must have
\begin{equation}
\sum_i A_i \geq N,
\label{areatest}
\end{equation}
where the sum ranges only over the beams which have not undergone caustics by
the given redshift.\footnote{There is an additional ``area test'' that 
should hold: If we add up {\em all} areas---with beams with an odd number
of caustics counting as negative and beams with an even number of 
caustics (or no caustics) counting as positive---we should (very nearly)
obtain the Robertson-Walker result. However, it was not possible to 
meaningfully apply this test to our Monte Carlo data, because
the total area in our data set 
was usually dominated by a single area value from a run which had
undergone a caustic (or a double-caustic). In other
words, our ``statistics'' were never adequate to investigate this test.} 

We have checked Eq.~(\ref{areatest}) in all of our Monte Carlo
simulations (including many not shown here) and have found it to be
satisfied in all cases. Furthermore, in all of our simulations in
which the matter distribution within ${\cal R}$ is spherically
symmetric (i.e., in all but the cylindrical cases),
the left side of Eq.~(\ref{areatest}) was larger than the right side
by only a tiny amount---typically, just a few percent.  This
remarkable (near) equality of the left and right sides of
Eq.~(\ref{areatest}) has two important consequences.  First,
it provides strong evidence of the self-consistency of our
calculations, since it is hard to imagine how such (near) equality
could hold for cases as different as, say, Figs.~\ref{F:z_3}
and~\ref{F:uni200}, if our Monte Carlo calculations were not properly
modeling at least some aspects of the spacetime
of Eq.~(\ref{metric}). Second, it shows that for spherical matter
distributions, almost all photons which leave the boundary of the past
of $p$ do so at (or very near to) a caustic. Consequently, it also
shows that in the spherical case, very few sources can have more than
one primary image.

For the case of a relatively ``thick'' cylinder (with negligible Weyl
curvature) as in Fig.~\ref{F:cyl52}, near equality also holds in
Eq.~(\ref{areatest}), as is evident from the fact that
Fig.~\ref{F:cyl52} is indistinguishable from
Fig.~\ref{F:uni200}. However, for a very thin cylinder as in
Fig.~\ref{F:cyl02}, the left side of Eq.~(\ref{areatest}) exceeds
the right side by a factor of about $1.8$. Thus, in
this case many sources must have multiple primary images, presumably
resulting from the passage of photons around different sides of the
cylinder.

\section{Correlations Between Quasar Luminosity and the Number of
Absorption Systems}
\label{york}
In a recent paper, Vanden Berk {\em et al.}~\cite{vandenberk} have
presented evidence for a positive correlation between quasar
luminosity and the number of intervening Carbon~IV absorption clouds.
Using the results of previous quasar surveys, these authors compiled a
catalog of nearly 500 quasars, with over 2,000 heavy metal
absorption lines. Analysis of this catalog revealed an excess of C~IV
absorbers in luminous QSOs. The authors proposed that this effect
might be caused by the brightening (i.e., magnification) of the quasar
images due to cumulative gravitational lensing by the mass distribution
associated with the C~IV absorbers.

Figs.~1 and~4 of \cite{vandenberk} present the main evidence in
favor of this positive correlation. In their Fig.~1, they divide the
quasars in their catalog into a ``bright half'' and a ``dim half'',
and they plot the number of C~IV absorbers for the two groups in five
different redshift bins. In all five bins, the bright quasars had more
C~IV absorbers than the dim ones. We have analyzed these results and
we estimate that if the effect found in their Fig.~1 were due
entirely to lensing, each C~IV absorber would have to contribute an
increase on the order of $1/2$~V~magnitude to the luminosity
of the quasar. Possible evidence for an even larger effect can be
found in Fig.~4 of \cite{vandenberk}. That figure shows a three
magnitude difference between the absolute luminosity of the quasars
with the largest and fewest number of C~IV absorbers. Taken at face
value, this suggests that each absorber would have to contribute an
increase of roughly $1$~V~magnitude to the luminosity of the quasar for
the effect to be due to lensing.

A proper analysis of the results of \cite{vandenberk} and their
implications for lensing would, of course, require a careful
consideration of numerous observational issues. We shall not attempt
to undertake such an analysis here.\footnote{The bias due to gravitational
lensing on the luminosity function of quasars has been investigated by
Pei~\cite{pei2}.}
Nevertheless, we shall pose the
following questions: Should gravitational lensing
produce a systematic, positive correlation between quasar luminosity
and the number of intervening absorption clouds? If so, is this effect
large enough to plausibly account for the results of
\cite{vandenberk}?

To analyze these questions, we make the extreme assumption that (most
of) the mass of the universe is associated with C~IV absorption
clouds. Clearly, this assumption should maximize the lensing effects
associated with the C~IV absorbers, and the results we obtain should
thereby be viewed as upper limits to the possible lensing effects of
these clouds. In our investigations, we considered underlying
Robertson-Walker models with $\Lambda = 0$ and with $\Omega_0$ either
$1$ or $0.1$. For each Robertson-Walker model, we considered a variety
of different possible mass distributions associated with the C~IV
absorbers. These mass distributions ranged from ``point masses''
(appropriate if the C~IV clouds are associated with large black holes
or with galaxies in which most of the matter has already condensed to
form stars) to low density isothermal balls (corresponding to galaxies
or proto-galaxies). For each cosmological model and mass distribution,
a ``cloud radius'' was defined so that in the Monte Carlo data we
generated, our average number of C~IV absorbers in the relevant
redshift range was equal (at least approximately) to the average
number of C~IV absorbers in the data of~\cite{vandenberk}. We then
performed the Monte Carlo calculations described in detail in the
previous sections above, but in addition, we now kept track of the
number of times the photon beam ``passed through a C~IV cloud'' (i.e.,
had an impact parameter smaller than the assigned ``cloud
radius''). At each redshift of interest, we again ordered our 2,000
``runs'' by area, and then binned these ordered runs in 100 groups
of 20 runs each. For each such group of 20, we then computed (from
our magnification data) the average magnitude of a ``standard candle''
source at the given redshift, as well as the average number of
``passages through C~IV clouds''. We then represented each of our
100 groups as a point in a ``scatter plot'' of average magnitude
versus average number of passages through a C~IV cloud.

\begin{figure}
\begin{center}
\begin{picture}(0,0)%
\epsfig{file=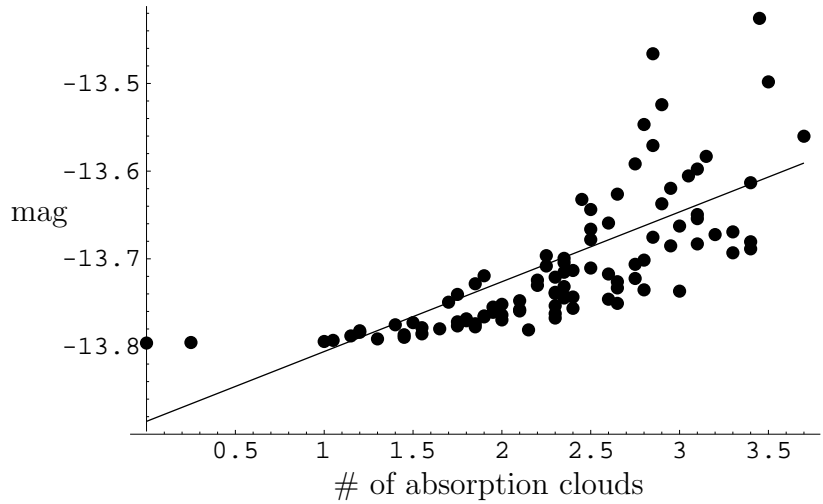}%
\end{picture}%
\setlength{\unitlength}{3947sp}%
\begingroup\makeatletter\ifx\SetFigFont\undefined%
\gdef\SetFigFont#1#2#3#4#5{%
  \reset@font\fontsize{#1}{#2pt}%
  \fontfamily{#3}\fontseries{#4}\fontshape{#5}%
  \selectfont}%
\fi\endgroup%
\begin{picture}(5097,3115)(3301,-4664)
\put(3301,-2911){\makebox(0,0)[lb]{\smash{\SetFigFont{12}{14.4}{\familydefault}{\mddefault}{\updefault}
mag
}}}
\put(5328,-4628){\makebox(0,0)[lb]{\smash{\SetFigFont{12}{14.4}{\familydefault}{\mddefault}{\updefault}
\# of absorption clouds
}}}
\end{picture}
\end{center}
\caption{Magnitude vs. \# of absorption clouds, for an $\Omega_0=0.1$,
$\Lambda=0$ universe, with galaxies treated as isothermal balls
of mass $5\times10^{11} M_\odot$ and radius $50\ \mbox{kpc}$. The
``magnitudes'' are for (standard candle) quasars at redshift $2$ (with
an arbitrarily chosen absolute luminosity). The ``number of absorption
clouds'' is the number of instances in which the photon beam passes
through a galaxy in the redshift range $1<z<2$. The data is for 2,000
runs, and has been binned in groups of $20$. The best fit line yields
a slope of $0.08$ magnitude per absorber.}
\label{F:york1}
\end{figure}

Fig.~\ref{F:york1} shows the results we obtained for (standard candle)
quasars at a redshift of $2$ for the case of a universe with
$\Omega_0=0.1$ and $\Lambda=0$, and with matter in isothermal balls of mass
$5\times10^{11} M_\odot$ and radius $50\ \mbox{kpc}$. In this case,
the ``cloud radius'' also was taken to be $50\ \mbox{kpc}$, so that
each time a photon beam punctures a galaxy, we say that we have
registered an absorption line.  It can be seen from the figure that
there is a clear positive correlation between luminosity and the
number of ``passages through clouds''. However, the effect is not very
large, corresponding to only about $0.1$ magnitude per absorber.

The results we obtained in a variety of other cases were quite
similar. In all the cases we examined, a positive correlation was
found between quasar luminosity and the number of ``passages through
clouds''. When we made the mass distribution associated with the
clouds more highly concentrated than in Fig.~\ref{F:york1}, the
lensing effect on luminosity became larger, but the correlation with
the number of clouds generally became less strong (since the lensing
effects were more dominated by a single close encounter), so the
overall results were not greatly changed. Thus, none of the cases we
investigated produced lensing effects which deviated significantly
from the roughly $0.1$ magnitude per absorber seen in
Fig.~\ref{F:york1}.  This value is undoubtedly an overestimate
since, as mentioned above, in our calculations we associated all of
the mass in the universe with the absorption clouds. Thus, although
our results definitely confirm that gravitational lensing should
produce a positive correlation between quasar luminosity and the
number of absorbers, it appears unlikely that the effects of
gravitational lensing are large enough to explain the results of
\cite{vandenberk}.

\section{Lensing ``Noise'' and Bias in Measurements of $q_0$ Using 
Supernova Luminosities}
\label{supernovae}

One of the key goals of observational cosmology is to determine the
parameters of what we have referred to as the underlying
Robertson-Walker model of our universe, particularly the deceleration
parameter, $q_0 = \Omega_0/2 - \Omega_\Lambda$ (where $\Omega_\Lambda
\equiv \Lambda/3H_0{}^2)$.  Much progress has been made recently in this
regard by using type Ia supernovae as standard
candles~\cite{perlmutter}.  The intrinsic dispersion of peak
magnitudes of type Ia supernovae is of the order $0.2$ magnitudes, and this
number has been steadily decreasing with improved understanding of the
phenomenology of type Ia supernova lightcurves.  An important issue is
whether or not lensing could produce a significant further dispersion
of (distant) supernovae magnitudes. If so, lensing could be an
important source of noise in estimates of $q_0$. In addition, since
the dispersion in luminosity produced by lensing is non-Gaussian and
can be highly asymmetric about the mean---with a high probability for a
small decrease in luminosity and a small probability for a large
increase---lensing also could produce a significant bias in the
results if there are selection effects or if the data is not properly
analyzed. Indeed, using a Swiss cheese model to investigate this
issue, Kantowski {\em et al.}~\cite{kantowski2} have claimed that
lensing can have a large effect, causing a bias as large as $33 \%$ in
$q_0$ measurements from supernovae at $z\sim0.5$ when the true
underlying Robertson-Walker model is one with $\Omega = 1$ and
$\Lambda = 0$. However, using other methods, Frieman has
recently argued that the induced flux dispersion in type Ia supernova
magnitudes due to lensing is less than about $0.04$ magnitudes for sources
within a redshift of $0.5$~\cite{frieman}. If so, lensing would not,
at present, be a significant source of noise, but might become
important if the current intrinsic spread in type Ia supernova
magnitudes can be halved through better understanding of the
phenomenology of the light curves. Frieman's estimates are consistent
with those of Wambsganss {\em et al.}~\cite{wambsganss}.

Our approach can be used to obtain the spread in image magnification
(and, hence, amplification) of a standard candle at any given
redshift, for any given cosmological parameters, and any choice of
inhomogeneous distribution of matter. According to our conjecture in
Subsection~\ref{ss:conjecture}, randomly distributed point masses
should provide the most noise and/or bias, so it is particularly
instructive to examine that case. Furthermore, as discussed in
Sections~\ref{method} and~\ref{points_and_clustering},
this case should provide a realistic description of
lensing phenomena in our universe if most of the matter in the
universe is clumped into stars.

Consider, first, the case of a universe with $\Omega = 1$ and $\Lambda
= 0$ filled with randomly distributed point masses. A plot of area
versus percentage of photon beams at $z = 0.5$ was previously given in
Figure~\ref{F:z_05}. We wish to convert this figure into a probability
distribution for the apparent luminosity of a ``standard candle''
source randomly placed on a sphere of radius $D$ centered on us,
corresponding to $z = 0.5$. As we argued in
Subsection~\ref{ss:results}, each source should have exactly one
primary image. It is straightforward to obtain the probability
distribution for the apparent luminosity of this primary image---and
we shall do so below. However, as discussed in
Subsection~\ref{ss:results}, we do not have a good way of determining
which secondary images are associated with a given primary image, so
we cannot directly obtain the probability distribution for the total
apparent luminosity associated with a source. This is not a very
serious problem in the present case, since Fig.~\ref{F:z_05} shows that
less than $5 \%$ of the photon beams (as measured in the ``present
sky'') have undergone a caustic by $z = 0.5$, so that less than $5 \%$
of the total expected luminosity of the sources at $z = 0.5$ will be
carried in secondary images. Undoubtedly, most of the luminosity
carried by the secondary images will be associated with sources whose
primary images are strongly lensed. Thus, if the primary and secondary
images of a source cannot be resolved (as would be the case for
microlensing by stars), the effect of including the secondary images
should be merely to further brighten a few of the sources with the
brightest primary images. Thus, the probability distribution we give
below for the apparent luminosity of the primary images should be
accurate for the total luminosity, except for the brightest sources.

\begin{figure}
\begin{center}
\begin{picture}(0,0)%
\epsfig{file=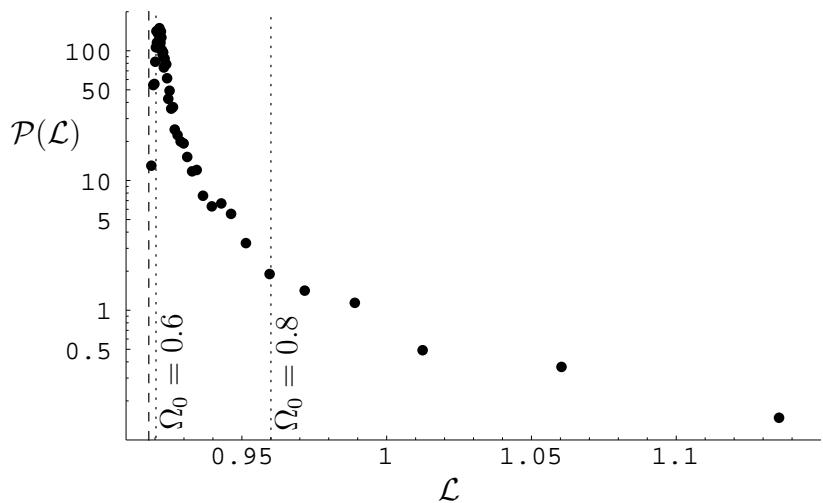}%
\end{picture}%
\setlength{\unitlength}{3947sp}%
\begingroup\makeatletter\ifx\SetFigFont\undefined%
\gdef\SetFigFont#1#2#3#4#5{%
  \reset@font\fontsize{#1}{#2pt}%
  \fontfamily{#3}\fontseries{#4}\fontshape{#5}%
  \selectfont}%
\fi\endgroup%
\begin{picture}(5099,3113)(3903,-4062)
\put(6587,-4023){\makebox(0,0)[lb]{\smash{\SetFigFont{12}{14.4}{\familydefault}{\mddefault}{\updefault}
${\cal L}$
}}}
\put(3903,-1780){\makebox(0,0)[lb]{\smash{\SetFigFont{12}{14.4}{\familydefault}{\mddefault}{\updefault}
${\cal P}({\cal L})$
}}}
\put(4980,-3588){\rotatebox{90.0}{\makebox(0,0)[lb]{\smash{\SetFigFont{12}{14.4}{\familydefault}{\mddefault}{\updefault}
$\Omega_0=0.6$
}}}}
\put(5701,-3596){\rotatebox{90.0}{\makebox(0,0)[lb]{\smash{\SetFigFont{12}{14.4}{\familydefault}{\mddefault}{\updefault}
$\Omega_0=0.8$
}}}}
\end{picture}
\end{center}
\caption{The probability distribution, ${\cal P}({\cal L})$, for the
apparent luminosity, ${\cal L}$, of a ``standard candle'' for an
$\Omega=1$, $\Lambda=0$ universe with point mass galaxies, at a
redshift of $1/2$.  The absolute luminosity of the standard candle has
been normalized to yield an apparent luminosity of $1$ in the
underlying Robertson-Walker model. The probability distribution shown
is for primary images only; inclusion of the flux from secondary
images presumably would mainly increase the luminosity of the
most luminous primary images (which are off the scale of this
plot), and should not significantly affect this figure.  (Note that,
according to Fig.~\ref{F:z_05}, approximately $5 \%$ of the total
luminosity is carried by secondary images.) The vertical dashed line
represents the empty beam apparent luminosity, which is the
minimum possible apparent luminosity for primary images. This empty
beam apparent luminosity corresponds to a Robertson-Walker model
with $\Omega_0=0.6$ and $\Lambda = 0$. The vertical dotted lines show
the lower and upper $16 \%$ of this probability distribution, to give
an indication of what one might roughly view as ``one sigma'' errors
in this highly non-Gaussian distribution with infinite second
moment. The $\Omega$ values corresponding to these lines also are
shown.}
\label{F:mu1}
\end{figure}

To convert Fig.~\ref{F:z_05} to a probability distribution for
apparent luminosity for the primary image of a randomly placed source,
we proceed as follows. Let $p_z(A)\,dA$ denote the probability that a
beam---which is randomly chosen with respect to the ``present
sky''---will have area between $A$ and $A+dA$ at redshift $z$. Up to
normalization, $p_{1/2}(A)$ is just the inverse of the slope of the
curve plotted in Figure~\ref{F:z_05}. Let $P_z(A)\,dA$ denote the
probability that a source which is randomly placed on a sphere
centered about us of radius $D$, corresponding to redshift $z$, will
be ``hit'' by a beam with area between $A$ and $A + dA$ which has not
undergone a caustic. Then, as previously mentioned in
Subsection~\ref{ss:results} above, we have
\begin{equation}
P_z(A) \propto A\,p_z(A).
\label{pP2}
\end{equation}
Since the apparent luminosity, ${\cal L}$, of the source is
proportional to $1/A$, the probability distribution, ${\cal P}_z({\cal
L})$, for apparent luminosity is given by
\begin{equation}
{\cal P}_z({\cal L})\propto {\cal L}^{-2} P_z(1/{\cal L}) 
\propto {\cal L}^{-3} p_z(1/{\cal L}),
\label{PP}
\end{equation}
where we have normalized both the beam area and intrinsic luminosity
so that both $A$ and ${\cal L}$ would have unit value at redshift $z$
in the underlying Robertson-Walker model. The probability
distribution, ${\cal P}_{1/2}({\cal L})$, is plotted in
Fig.~\ref{F:mu1}, using the data from Fig.~\ref{F:z_05} to
determine $p_{1/2}(A)$.

It should be noted that, since at any $z$ we have $p_z(0) \neq 0$, it
follows from Eq.~(\ref{PP}) that as ${\cal L} \rightarrow \infty$, we
have at all $z$
\begin{equation}
{\cal P}({\cal L}) \propto 1/{\cal L}^3.
\label{PL}
\end{equation}
Consequently, ${\cal P}({\cal L})$ is normalizable (as it must be) and
has a well defined first moment (since, as mentioned in
Subsec.~\ref{ss:checks}, the total expected apparent luminosity
[including the secondary as well as primary images] must agree with
that of the underlying Robertson-Walker model). However, its second moment
is logarithmically divergent. As a result, the law of large numbers
fails to apply to ${\cal P}({\cal L})$. Thus, if one makes repeated
measurements (by observing many supernova events) and averages the
apparent luminosities, one will {\em not} obtain a sharply peaked
Gaussian distribution about the average value. If the strongly lensed
events are removed from the data sample, a Gaussian distribution would
be obtained for the average value, but a bias will be introduced
toward smaller apparent luminosity.

To obtain a rough, quantitative measure of both the ``noise'' and the
potential bias in apparent luminosity measurements resulting from
lensing, we inserted dotted lines in Fig.~\ref{F:mu1} to demarcate the
upper and lower $16 \%$ of the probability distribution. If the
probability distribution were a Gaussian, these lines would correspond
to a ``one sigma'' error centered about the mean. Thus, the separation
of these dotted lines gives a rough indication of the lensing
``noise'', whereas the ``off-centerness'' of the lines (away from $1$)
gives a rough indication of the potential bias that would occur if the
strong lensing events were not included in the data sample. We see
that in the point mass case, if the strong lensing events are
excluded, the potential exists for a significant bias toward values of
apparent luminosity nearly as small as the empty beam value. This
result is consistent with the results of Kantowski {\em et
al.}~\cite{kantowski2}, since the imposition of an ``opaque radius''
effectively excises the strong lensing events.

\begin{figure}
\begin{center}
\begin{picture}(0,0)%
\epsfig{file=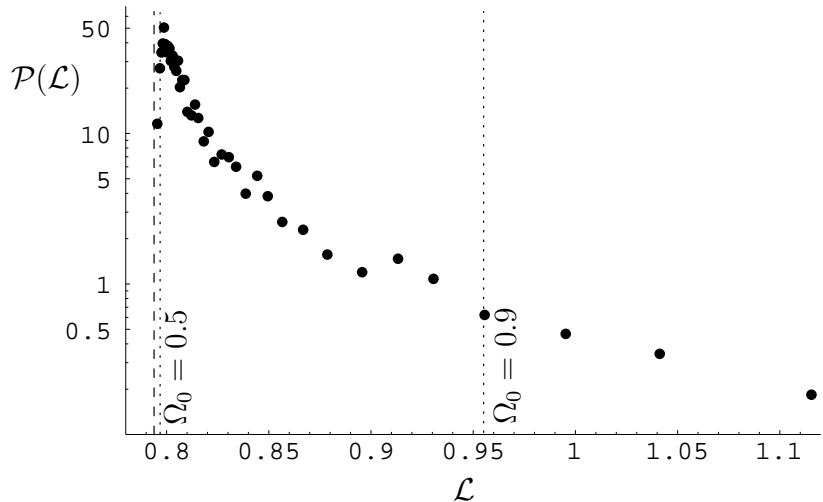}%
\end{picture}%
\setlength{\unitlength}{3947sp}%
\begingroup\makeatletter\ifx\SetFigFont\undefined%
\gdef\SetFigFont#1#2#3#4#5{%
  \reset@font\fontsize{#1}{#2pt}%
  \fontfamily{#3}\fontseries{#4}\fontshape{#5}%
  \selectfont}%
\fi\endgroup%
\begin{picture}(5096,3149)(3902,-3498)
\put(6679,-3465){\makebox(0,0)[lb]{\smash{\SetFigFont{12}{14.4}{\familydefault}{\mddefault}{\updefault}
$\cal L$
}}}
\put(3902,-873){\makebox(0,0)[lb]{\smash{\SetFigFont{12}{14.4}{\familydefault}{\mddefault}{\updefault}
${\cal P}({\cal L})$
}}}
\put(5003,-2995){\rotatebox{90.0}{\makebox(0,0)[lb]{\smash{\SetFigFont{12}{14.4}{\familydefault}{\mddefault}{\updefault}
$\Omega_0=0.5$
}}}}
\put(7072,-2987){\rotatebox{90.0}{\makebox(0,0)[lb]{\smash{\SetFigFont{12}{14.4}{\familydefault}{\mddefault}{\updefault}
$\Omega_0=0.9$
}}}}
\end{picture}
\end{center}
\caption{The probability distribution for apparent luminosity, ${\cal
L}$, of a ``standard candle'' for an $\Omega=1$, $\Lambda=0$
universe with point mass galaxies, at a redshift of $1$. The dashed
and dotted lines have the same meaning as in
Figure~\ref{F:mu1}. Again, only primary images are considered. In this
case, about $12 \%$ of the total luminosity is carried by secondary
images (see Figure~\ref{F:z_1}), so the corrections to the plot which
would result from inclusion of secondary images may be somewhat more
significant than in Figure~\ref{F:mu1}.}
\label{F:mu1a}
\end{figure}

The corresponding plot at $z=1$ for a universe with $\Omega = 1$ and
$\Lambda = 0$ is shown in Fig.~\ref{F:mu1a}. As can be seen from this
figure, both the noise and potential bias due to gravitational lensing
are considerably larger at $z = 1$ than at $z = 1/2$.
Figs.~\ref{F:mu1} and~\ref{F:mu1a} appear to be in good qualitative agreement
with the results of~\cite{rauch,pei1}.

\begin{figure}
\begin{center}
\begin{picture}(0,0)%
\epsfig{file=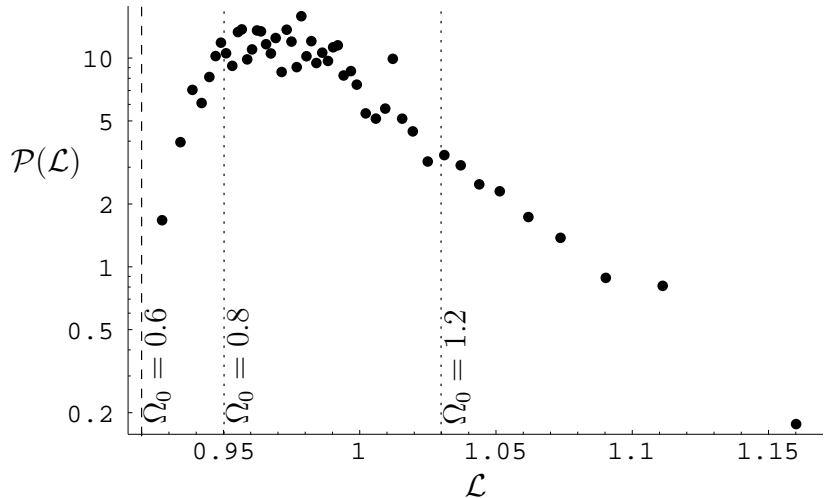}%
\end{picture}%
\setlength{\unitlength}{3947sp}%
\begingroup\makeatletter\ifx\SetFigFont\undefined%
\gdef\SetFigFont#1#2#3#4#5{%
  \reset@font\fontsize{#1}{#2pt}%
  \fontfamily{#3}\fontseries{#4}\fontshape{#5}%
  \selectfont}%
\fi\endgroup%
\begin{picture}(5100,3131)(5098,-4080)
\put(7951,-4041){\makebox(0,0)[lb]{\smash{\SetFigFont{12}{14.4}{\familydefault}{\mddefault}{\updefault}
${\cal L}$
}}}
\put(5098,-1992){\makebox(0,0)[lb]{\smash{\SetFigFont{12}{14.4}{\familydefault}{\mddefault}{\updefault}
${\cal P}({\cal L})$
}}}
\put(6073,-3587){\rotatebox{90.0}{\makebox(0,0)[lb]{\smash{\SetFigFont{12}{14.4}{\familydefault}{\mddefault}{\updefault}
$\Omega_0=0.6$
}}}}
\put(6589,-3587){\rotatebox{90.0}{\makebox(0,0)[lb]{\smash{\SetFigFont{12}{14.4}{\familydefault}{\mddefault}{\updefault}
$\Omega_0=0.8$
}}}}
\put(7954,-3595){\rotatebox{90.0}{\makebox(0,0)[lb]{\smash{\SetFigFont{12}{14.4}{\familydefault}{\mddefault}{\updefault}
$\Omega_0=1.2$
}}}}
\end{picture}
\end{center}
\caption{The probability distribution for apparent luminosity, ${\cal
L}$, of a ``standard candle'' for an $\Omega=1$, $\Lambda=0$
universe with isothermal galaxies of radius $200\ \mbox{kpc}$, at a
redshift of $1/2$.}
\label{F:mu2}
\end{figure}

\begin{figure}
\begin{center}
\begin{picture}(0,0)%
\epsfig{file=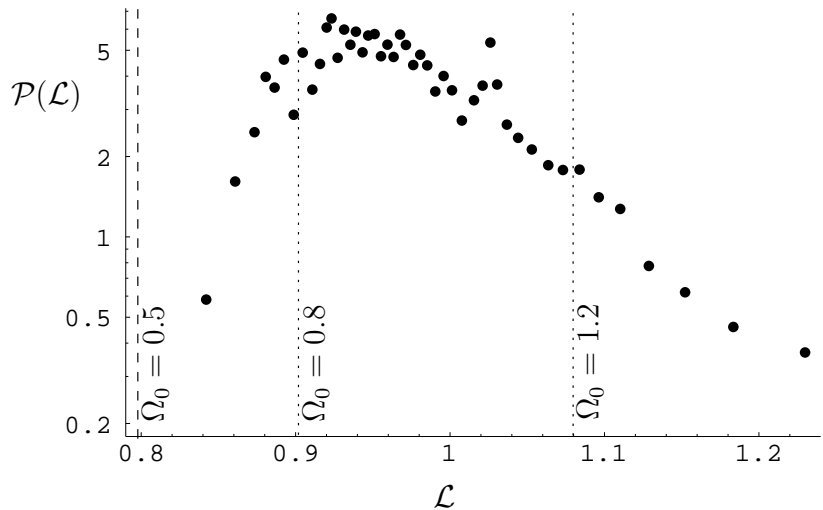}%
\end{picture}%
\setlength{\unitlength}{3947sp}%
\begingroup\makeatletter\ifx\SetFigFont\undefined%
\gdef\SetFigFont#1#2#3#4#5{%
  \reset@font\fontsize{#1}{#2pt}%
  \fontfamily{#3}\fontseries{#4}\fontshape{#5}%
  \selectfont}%
\fi\endgroup%
\begin{picture}(5095,3195)(5103,-4144)
\put(7757,-4105){\makebox(0,0)[lb]{\smash{\SetFigFont{12}{14.4}{\familydefault}{\mddefault}{\updefault}
${\cal L}$
}}}
\put(5103,-1577){\makebox(0,0)[lb]{\smash{\SetFigFont{12}{14.4}{\familydefault}{\mddefault}{\updefault}
${\cal P}({\cal L})$
}}}
\put(6063,-3570){\rotatebox{90.0}{\makebox(0,0)[lb]{\smash{\SetFigFont{12}{14.4}{\familydefault}{\mddefault}{\updefault}
$\Omega_0=0.5$
}}}}
\put(7072,-3570){\rotatebox{90.0}{\makebox(0,0)[lb]{\smash{\SetFigFont{12}{14.4}{\familydefault}{\mddefault}{\updefault}
$\Omega_0=0.8$
}}}}
\put(8794,-3563){\rotatebox{90.0}{\makebox(0,0)[lb]{\smash{\SetFigFont{12}{14.4}{\familydefault}{\mddefault}{\updefault}
$\Omega_0=1.2$
}}}}
\end{picture}
\end{center}
\caption{The probability distribution for apparent luminosity, ${\cal
L}$, of a ``standard candle'' for an $\Omega=1$, $\Lambda=0$
universe with isothermal galaxies of radius $200\ \mbox{kpc}$, at a
redshift of $1$.}
\label{F:mu3}
\end{figure}

The noise and potential bias due to lensing are considerably smaller
if microlensing is not taken into account, as would be justified if
most of the matter is smoothly distributed, rather than clumped into
stars. As an example of a model with smoothly distributed matter,
consider, again, the case of a universe with $\Omega = 1$ and $\Lambda
= 0$, but now with the galaxies treated as isothermal balls of radius
$200\ \mbox{kpc}$.  The probability distribution ${\cal P}({\cal L})$
for redshifts of~$1/2$ and~$1$ are shown in Figs.~\ref{F:mu2}
and~\ref{F:mu3}. There are very few caustics even at $z = 1$ in this
case, so the secondary images are of no importance. Inspection of
Figs.~\ref{F:mu2} and~\ref{F:mu3} shows that the probability
distribution is peaked much closer to $1$ than in the point mass case,
and the high luminosity ``tail'' of the distribution is much
smaller. Thus, there would appear to be no significant danger of
``bias'' in this case. In addition, there is considerably less
``noise'' than in the point mass case. The noise estimate obtained
from the dotted lines in Fig.~\ref{F:mu2} is in good agreement with
the estimate obtained by Frieman~\cite{frieman}.

\begin{figure}
\begin{center}
\begin{picture}(0,0)%
\epsfig{file=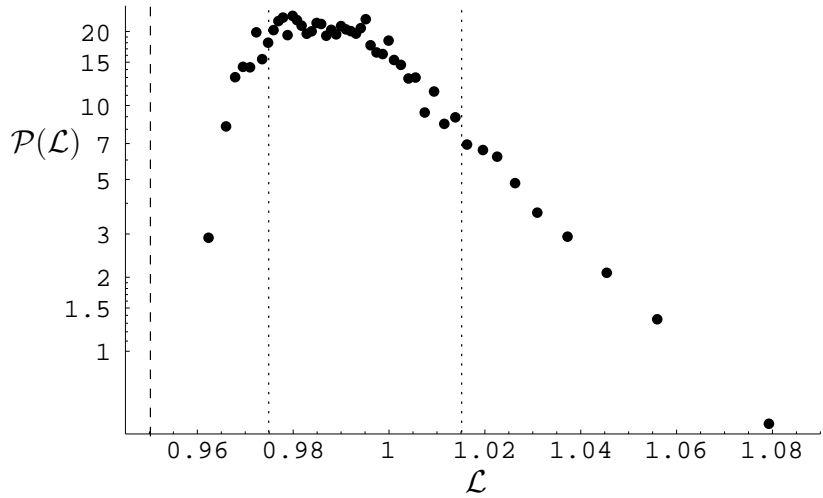}%
\end{picture}%
\setlength{\unitlength}{3947sp}%
\begingroup\makeatletter\ifx\SetFigFont\undefined%
\gdef\SetFigFont#1#2#3#4#5{%
  \reset@font\fontsize{#1}{#2pt}%
  \fontfamily{#3}\fontseries{#4}\fontshape{#5}%
  \selectfont}%
\fi\endgroup%
\begin{picture}(5097,3101)(5701,-4050)
\put(5701,-1881){\makebox(0,0)[lb]{\smash{\SetFigFont{12}{14.4}{\familydefault}{\mddefault}{\updefault}
${\cal P}({\cal L})$
}}}
\put(8554,-4017){\makebox(0,0)[lb]{\smash{\SetFigFont{12}{14.4}{\familydefault}{\mddefault}{\updefault}
$\cal L$
}}}
\end{picture}
\end{center}
\caption{The probability distribution for apparent luminosity, ${\cal
L}$, of a ``standard candle'' for an $\Omega_0=0.4$,
$\Omega_\Lambda=0.6$ universe with isothermal galaxies of radius
$200\ \mbox{kpc}$, at a redshift of $1/2$.}
\label{F:mu4}
\end{figure}

\begin{figure}
\begin{center}
\begin{picture}(0,0)%
\epsfig{file=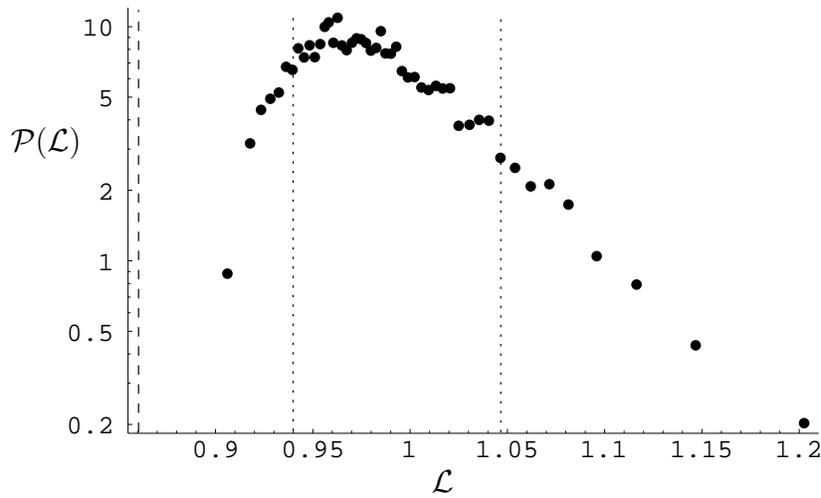}%
\end{picture}%
\setlength{\unitlength}{3947sp}%
\begingroup\makeatletter\ifx\SetFigFont\undefined%
\gdef\SetFigFont#1#2#3#4#5{%
  \reset@font\fontsize{#1}{#2pt}%
  \fontfamily{#3}\fontseries{#4}\fontshape{#5}%
  \selectfont}%
\fi\endgroup%
\begin{picture}(5100,3083)(5698,-4032)
\put(5698,-1854){\makebox(0,0)[lb]{\smash{\SetFigFont{12}{14.4}{\familydefault}{\mddefault}{\updefault}
${\cal P}({\cal L})$
}}}
\put(8343,-3999){\makebox(0,0)[lb]{\smash{\SetFigFont{12}{14.4}{\familydefault}{\mddefault}{\updefault}
$\cal L$
}}}
\end{picture}
\end{center}
\caption{The probability distribution for apparent luminosity, ${\cal
L}$, of a ``standard candle'' for an $\Omega_0=0.4$,
$\Omega_\Lambda=0.6$ universe, with isothermal galaxies of radius
$200\ \mbox{kpc}$, at a redshift of $1$.}
\label{F:mu5}
\end{figure}

Wambsganss {\em et al.}~\cite{wambsganss} have investigated a
cosmological model with $\Omega_0=0.4$ and $\Omega_\Lambda=0.6$ using
a cold dark matter N-body computer simulation (normalized to COBE
data) to determine the matter distribution. They studied lensing in
this model and produced plots of ${\cal P}({\cal L})$ at $z = 1/2$ and
$z = 1$. It is instructive to compare their results with what would be
obtained from our approach. To do so, we considered a $\Omega_0=0.4$,
$\Omega_\Lambda=0.6$ universe and distributed the matter in galaxies
chosen to be isothermal balls of radius $200\ \mbox{kpc}$. (This
choice of galactic mass distribution is merely our guess as to what
should be reasonable for this cosmology; we did not attempt to match
the results of the simulation used in~\cite{wambsganss}.)  In order to
obtain good statistics for the comparison, we did 6,000 (rather than
our usual 2,000) runs in this case. Our results for ${\cal P}({\cal
L})$ at $z = 1/2$ and $z = 1$ are shown in Figs.~\ref{F:mu4}
and~\ref{F:mu5} respectively. Comparison with Fig.~1
of~\cite{wambsganss} shows very good agreement---apart from an
overall normalization (which appears to have been chosen arbitrarily
in~\cite{wambsganss}). Indeed, this agreement seems remarkably good in
view of the fact that we did not attempt to adjust our galactic mass
distribution to theirs, and, in our calculations, the clustering of
galaxies is {\em not} taken into account. This latter fact lends
support to the argument presented in
Subsection~\ref{ss:relevant_scales} that, in almost all models,
clustering of galaxies should have a negligible effect on the
statistical distributions for magnification, shear, and rotation.

Finally, we note that a knowledge of ${\cal P}({\cal L})$ over a range
of redshifts contains a great deal of information about both the
nature of the mass distribution in the universe and the parameters of
the underlying Robertson-Walker cosmology. In particular, the shape of
${\cal P}({\cal L})$ can be used to determine the fraction of matter
in the form of condensed objects. We emphasize that, to do so, it
should not be necessary to go to the high luminosity tail of the
distribution, as has been considered previously~\cite{rauch,wagoner},
since there are considerable differences between, e.g.,
Figs.~\ref{F:mu1a} and~\ref{F:mu3} at low apparent luminosities as
well.

With regard to determining the parameters of the underlying
Robertson-Walker cosmology, as indicated above, due
to photon conservation the expected total luminosity---including both
primary and secondary images---of all sources is independent of
gravitational lensing effects. Thus, if no other sources of noise or
bias are present, the apparent luminosity of a standard candle in the
underlying Robertson-Walker model always can be obtained, in
principle, by averaging the apparent luminosities of the observed
sources. We stress that, since the probability distribution, ${\cal
P}({\cal L})$, can be quite asymmetric about its mean, it is crucial
that one averages apparent luminosities rather than some function of
them (like apparent magnitudes). Furthermore, in the point mass
case---relevant if microlensing by stars is important---it will be
necessary to make efforts to include the very bright images in the
average in order not to bias the results toward smaller apparent
luminosity.

\section*{Acknowledgements}

The main stimulus for this work came from Don York, Dan Vanden Berk,
and Jean Quashnock, who posed to us the issue of whether the results
found in \cite{vandenberk} could be accounted for by gravitational
lensing. We wish to thank them for providing this initial stimulus and
for many subsequent fruitful discussions. We also wish to thank
J\"urgen Ehlers for reading a preliminary version of this manuscript,
and \'Eanna Flanagan and Josh Frieman for reading (nearly) final
versions. Finally we wish to thank Eric Linder for giving a
careful reading of the paper, and for many valuable comments.
Our research was supported by NSF grant PHY 95-14726 to the
University of Chicago.

\end{document}